\documentclass[aip,jcp,preprint, notitlepage, citeautoscript]{revtex4-2}

\usepackage[version=3]{mhchem} % Formula subscripts using \ce{}
\usepackage{braket}
\usepackage{mathtools}
\usepackage{algorithm}
\usepackage{algorithmic}
\usepackage{geometry}
\usepackage{amsmath}
\usepackage{amsthm}
\usepackage{graphicx}
\usepackage{fancyvrb}
\usepackage{forest}
\usepackage{tikz-qtree}
\usepackage[dvipsnames]{xcolor}
\usepackage[english]{babel} 
\makeatletter
\def\frontmatter@preabstractspace{0pt}
\def\frontmatter@postabstractspace{0pt}

\def\frontmatter@abstractfont{\footnotesize}
\makeatother
\begin{document}
\title{Initialization with a Fock State Cavity Mode in Real-Time Nuclear--Electronic Orbital Polariton Dynamics}
\author{Millan F. Welman}
\author{Sharon Hammes-Schiffer}
\affiliation{Department of Chemistry, Princeton University, Princeton, NJ 08544}
%\date{\today}
\begin{abstract}
Molecular polaritons have drawn great interest in recent years as a possible avenue for providing optical control over chemical dynamics. A central challenge in the field is to identify physical phenomena that require a quantum rather than a classical treatment of electrodynamics. In this work, we use our recently developed mean-field quantum (mfq) and full-quantum (fq) real-time nuclear--electronic orbital (RT-NEO) time-dependent density functional theory methods to simulate polaritonic dynamics for a molecule under vibrational strong coupling when a quantized cavity mode is initialized in a Fock state rather than a coherent state. Our previous work showed that a coherent state initial condition for the cavity mode leads to polariton formation for both the mfq-RT-NEO and fq-RT-NEO methods. Herein, we show that the mfq-RT-NEO method, which does not allow light--matter entanglement, does not predict polariton formation for a Fock state initial condition. Similar to the mfq-RT-NEO method, the fq-RT-NEO method does not predict oscillations of the cavity mode coordinate and molecular dipole operator expectation values for a Fock state initial condition. However, the fq-RT-NEO method does predict oscillations of the expectation values of even powers of these operators as well as light--matter entanglement, implicating polariton formation with a Fock state initial condition. All these observations can be explained with model systems. These results suggest that using a quantized cavity mode initial condition that does not have a direct analogy to an initial condition in classical electrodynamics can lead to physical phenomena that can only be described by a quantum treatment of the cavity mode.
\end{abstract}
\maketitle

%%%%%%%%%%%%%%%%%%%%%%%%%%%%%%%%%%%%%%%%%%%%%%%%%%%%%%%%%%%%%%%%%%%%%
%% Start the main part of the manuscript here.
%%%%%%%%%%%%%%%%%%%%%%%%%%%%%%%%%%%%%%%%%%%%%%%%%%%%%%%%%%%%%%%%%%%%%
\section{Introduction}
Molecular polaritons are quasiparticles formed by strong coupling between light and matter degrees of freedom\cite{mandal_theoretical_2023}. When an incident electromagnetic field interacts with a system of one or more molecules, it induces an oscillation in the charge distribution of the molecular system. The oscillating molecular charge distribution in turn generates another electromagnetic field, which can interact with and alter the behavior of the incident field that created it. This feedback cycle will occur if the dissipation timescale of the incident field is comparable to or much slower than the timescale of energy exchange between the molecular system and the electromagnetic field\cite{xiong_molecular_2023}. This phenomenon can be realized experimentally in settings such as plasmonic nanocavities \cite{pelton_strong_2019}, where a small number of molecules interacts with a plasmonic mode, or Fabry-P\'erot optical cavities\cite{xiang_molecular_2024, li_molecular_2022}, where a large number of molecules is confined between two mirrors and allowed to interact with cavity modes bounded by the mirrors. The total light--matter system is said to be under electronic strong coupling (ESC)\cite{deng_exciton-polariton_2010, keeling_boseeinstein_2020} when the electromagnetic mode is coupled to an electronic transition and under vibrational strong coupling (VSC) \cite{long_coherent_2015, george_liquid-phase_2015, wright_rovibrational_2023} when the electromagnetic mode is coupled to a vibrational transition. 

Recent experimental data suggest that strong light--matter coupling may be used to modify chemical dynamics ranging from energy transfer and exciton transport processes\cite{zhong_energy_2017, xiang_intermolecular_2020, garcia-vidal_manipulating_2021, georgiou_ultralongrange_2021, kertzscher_tunable_2024, lidzey_strong_1998, feist_extraordinary_2015, schachenmayer_cavity-enhanced_2015, coles_polariton-mediated_2014, kasprzak_boseeinstein_2006} to reaction rates \cite{ahn_modification_2023, thomas_ground_2020, lather_cavity_2019} and branching ratios \cite{thomas_tilting_2019}. This experimental progress in turn has spurred the development of a variety of theoretical methods ranging from canonical quantum optics models \cite{jaynes_comparison_1963, tavis_exact_1968, tavis_approximate_1969} to quantum-electrodynamical (QED) electronic structure methods \cite{schafer_ab_2018, deprince_cavity-modulated_2021, mallory_reduced-density-matrix-based_2022, vu_cavity_2024, el_moutaoukal_strong_2025, bauer_perturbation_2023, mctague_non-hermitian_2022, pavosevic_wavefunction_2022, cui_variational_2024}, which have shed light on the nature of polaritonic stationary states and the dynamics of polaritonic model systems. 
Among theoretical methods for simulating molecular polaritons, dynamical methods provide a unique perspective by capturing the feedback between light and matter in real time. An important consideration is whether the electromagnetic field should be treated classically or quantum mechanically. Either choice can be used to simulate polariton dynamics because the feedback between light and matter is predicted by both classical and quantum electrodynamics. Each approach has its own distinct advantages. 

Semiclassical treatments of light--matter interactions, where a classical electromagnetic field is coupled to a quantum mechanical matter system, have found widespread applications in the study of problems ranging from weak to strong light--matter coupling \cite{cotton_symmetrical_2013, miller_semiclassical_2001, sukharev_optics_2017, bustamante_dissipative_2021, tancogne-dejean_octopus_2020, yamada_time-dependent_2018, chen_classical_2010, chen_predictive_2019}. Indeed, a semiclassical method often constitutes the only computationally feasible approach to simulating the collective strong light--matter coupling that occurs experimentally in a Fabry-P\'erot cavity. The cavity molecular dynamics (CavMD) method \cite{li_cavity_2020} provides such a framework for carrying out computationally tractable simulations of collective strong coupling under VSC. CavMD simulations have thus far yielded significant insights into relaxation and energy transfer processes in non-reactive dynamics under VSC \cite{li_cavity_2021, li_collective_2021, li_polariton_2022, li_energy-efficient_2022}.  In addition to the extensive theoretical literature employing semiclassical methods, it is well-known experimentally that classical transfer matrix methods accurately predict polaritonic absorption lineshapes in the limit of a large molecular ensemble, as would occur in a Fabry-P\'erot cavity \cite{schwennicke_when_2024}. This result has recently been verified analytically \cite{yuen-zhou_linear_2024}, suggesting that a semiclassical treatment may provide an acceptable description of light--matter interactions under collective strong coupling. 

In contrast to the semiclassical treatment, a fully quantum mechanical approach that features a quantized electromagnetic field allows for the inclusion of effects such as light--matter entanglement and spontaneous emission. Such a treatment may be computationally feasible for a system composed of a few molecules, as realized experimentally in a plasmonic nanocavity \cite{li_molecular_2022, xiang_molecular_2024}. Further motivation for a fully quantum mechanical treatment is provided by recent model system studies suggesting that there may be limitations to the semiclassical treatment of polariton dynamics \cite{michael_ruggenthaler_quantum_2014, flick_kohnsham_2015, flick_atoms_2017, ke_quantum_2024, fiechter_how_2023, simko_twin_2025, zeb_exact_2018, koner_hidden_2025}. Therefore, it is of great importance to better understand the potential limitations of a semiclassical treatment relative to a fully quantum mechanical treatment and how those limitations may manifest in dynamical simulations of polaritons.

To address this challenge, we recently developed a hierarchy of first-principles dynamical simulation methods for molecular polaritons. \cite{li_semiclassical_2022, welman_light-matter_2025} These methods treat the molecular system with real-time nuclear--electronic orbital time-dependent density functional theory (RT-NEO-TDDFT, hereafter RT-NEO) \cite{zhao_real-time_2020}. The RT-NEO method is the extension of conventional electronic RT-TDDFT to the NEO framework. In the NEO framework,\cite{webb_multiconfigurational_2002, hammes-schiffer_nuclearelectronic_2021} specified nuclei, usually protons, are treated quantum mechanically on the same footing as the electronic degrees of freedom. The RT-NEO method provides an approach for the propagation of nonequilibrium molecular quantum dynamics that incorporates zero-point energy, tunneling, proton delocalization, and other nuclear quantum effects. Coupling molecular dynamics propagated with RT-NEO to the dynamics of a cavity mode thus provides a first-principles route to simulating polariton dynamics under VSC that incorporates nuclear quantum effects. In this work, we will utilize two of the three methods in this hierarchy, omitting the semiclassical method, which treats the cavity mode classically.

Both RT-NEO methods that we will use in this work treat the cavity mode quantum mechanically but differ in their treatment of light--matter entanglement. The mean-field-quantum RT-NEO (mfq-RT-NEO) method self-consistently propagates separate equations of motion for the molecular density matrix and the cavity mode density matrix. The two density matrices are coupled to each other through the expectation values of the cavity mode coordinate operator and the molecular dipole operator. This treatment is equivalent to assuming that there is no quantum entanglement between the molecular system and the cavity mode. In contrast to the mfq-RT-NEO method, the full-quantum RT-NEO (fq-RT-NEO) method allows for light--matter entanglement by propagating a joint light-matter density matrix that describes both the molecular and cavity mode degrees of freedom. In Ref. \citenum{welman_light-matter_2025}, we used these methods to study non-reactive single-molecule polariton dynamics for the HCN molecule under vibrational strong coupling, where all electrons and the hydrogen nucleus are treated quantum mechanically. Under a consistent set of initial conditions, these two methods yielded excellent agreement in their description of the oscillations of the quantum nuclear dipole moment. The dynamics of the light--matter entanglement computed with the fq-RT-NEO method, however, revealed a new oscillation timescale absent from the mfq-RT-NEO results. This finding suggested that new physics might be uncovered by considering the role of light--matter entanglement.

The mfq-RT-NEO and fq-RT-NEO simulations in our previous work used a common set of initial conditions. The HCN molecular system was initialized in its ground state, while the cavity mode was initialized in a coherent state with a nonzero coordinate expectation value. This choice was made because coherent states provide an ideal initial condition for methods with a quantum cavity mode to enable a direct comparison to the results of semiclassical methods with a classical cavity mode. In the absence of light--matter coupling, the expectation values of the coordinate and momentum operators of a quantum cavity mode initialized in a coherent state will oscillate harmonically like a classical cavity mode. It is straightforward, however, to choose initial quantum mode conditions that do not lead to classical-like oscillations in the absence of light--matter coupling. The most obvious example of this is a Fock (number) state, or a quantum harmonic oscillator eigenstate, where the number of the state corresponds to the integer number of photons populating the cavity mode. This initial condition is of interest because it corresponds to the traditional model understanding of polaritons, where the cavity mode is populated with a single photon, and the photon is then exchanged coherently with a molecular system. Unlike a classical harmonic oscillator, a quantum harmonic oscillator in a Fock state can simultaneously have a nonzero energy expectation value and zero position and momentum expectation values. \cite{operator_note}. This insight suggests that more significant deviations from the general qualitative features of semiclassical polariton dynamics may be found if the quantum mode is  initialized in a Fock state. 

The goal of this paper is to perform mfq-RT-NEO and fq-RT-NEO simulations of single-molecule polariton dynamics with a quantum cavity mode initialized in a Fock state and to determine if these dynamics exhibit  qualitatively new features that do not occur in polariton dynamics where the cavity mode is treated classically or quantum mechanically with a coherent state initial condition. The paper is organized as follows. In section 2, we review the theory of mfq-RT-NEO and fq-RT-NEO, as well as our procedure for computing light--matter entanglement. Simulation details for HCN under vibrational strong coupling are provided in section 3. In section 4, we present the results of the mfq-RT-NEO and fq-RT-NEO calculations. Finally, section 5 provides a more detailed analysis of our results, aided by comparison to quantum optics models. Conclusions are provided in section 6.

\section{Theory}
In this section, we provide an overview of the mfq-RT-NEO and fq-RT-NEO methods. The interested reader is referred to Ref. \citenum{welman_light-matter_2025} for a more detailed derivation and discussion of these methods. 

Our starting point is the QED Hamiltonian written in the long-wavelength approximation:
\begin{equation}
    \hat{H}_{\rm QED} = \hat{H}_{\rm M} + \sum_{k,\lambda}\left[\frac{1}{2}\hat{p}_{k,\lambda}^2 + \frac{1}{2}\omega_{k,\lambda}^2\left( \hat{q}_{k,\lambda} + \frac{\boldsymbol{\hat{\mu}}_{\rm M}\cdot\boldsymbol{\xi}_\lambda}{\omega_{k,\lambda}\sqrt{\Omega\epsilon_0}}\right)^2\right]
    \label{qed hamiltonian}
\end{equation}
where $\hat{H}_{\rm M}$ is the molecular Hamiltonian,
\begin{equation}
    \hat{H}_{\rm M} = \sum_i\left(\frac{\hat{\boldsymbol{\rm p}}^2_i}{2m_i} + \hat{V}(\{\hat{\boldsymbol{\rm r}}_i\}, \{\textbf{R}_{\rm c}\})\right)
    \label{molecular hamiltonian}
\end{equation}
The summation with index $i$ runs over all quantum mechanical particles (i.e., electrons and specified quantum nuclei for NEO methods). The operator $\hat{V}(\{\hat{\boldsymbol{\rm r}}_i\}, \{\textbf{R}_{\rm c}\})$ includes all Coulomb interactions among quantum particles with coordinate operators $\{\hat{\boldsymbol{\rm r}}_i\}$ and classical nuclei at fixed coordinates $\{\textbf{R}_{\rm c}\}$. The summation in Eq. \ref{qed hamiltonian} is over all cavity modes with wave vector magnitude $k = |\boldsymbol{\rm k}| = \omega_{k, \lambda}/c$, where $c$ is the speed of light, and polarization direction given by the unit vector $\boldsymbol{\xi}_\lambda$. These two vectors are orthogonal, such that $\boldsymbol{\rm k}  \cdot \boldsymbol{\xi}_\lambda = 0$. The operators $\hat{q}_{k,\lambda}$ and $\hat{p}_{k,\lambda}$ and the scalar $\omega_{k,\lambda}$ are the coordinate operator, momentum operator, and frequency, respectively, of a mode determined by the vectors $\boldsymbol{\rm k}$ and $\boldsymbol{\xi}_\lambda$. The coordinate operator $\hat{q}_{k,\lambda}$ is coupled to the total molecular dipole moment $\boldsymbol{\hat{\mu}}_{\rm M}$. Finally, $\epsilon_0$ is the permittivity of free space, and $\Omega$ denotes the effective volume of the cavity. 

If we restrict our consideration to the strong coupling regime, avoiding the ultrastrong coupling regime, we can neglect the term in Eq. 1 that is quadratic in $\boldsymbol{\hat{\mu}}_{\rm M}$. We also assume that the total molecular density matrix $\textbf{P}_{\rm en}(t)$ can be written as a product of separate electronic and quantum nuclear density matrices:
\begin{equation}
    \textbf{P}_{\rm en}(t) = \textbf{P}_{\rm e}(t) \otimes \textbf{P}_{\rm n}(t)
    \label{molecular separability ansatz}
\end{equation}

Introducing the subscript F (for ``field") to denote quantities defined in the cavity mode Hilbert space, we can express the fq-RT-NEO equations of motion for the joint density matrix $\textbf{P}_{\rm Fn}(t)$ describing the cavity mode and nuclear degrees of freedom and the density matrix $\textbf{P}_{\rm e}(t)$ describing the electronic degrees of freedom as
\begin{equation}
    i\hbar\frac{\partial}{\partial t}\textbf{P}_{\rm Fn}(t) = \left[\textbf{I}_{\rm F} \otimes \textbf{F}^{\rm{NEO}}_{\rm n}(t) + \textbf{H}_{{\rm F}} \otimes \textbf{I}_{\rm n} + \sum_{k,\lambda}\varepsilon_{k,\lambda} \textbf{q}_{k,\lambda}\otimes\left(\boldsymbol{\mu}_{{\rm n}, \lambda} - \mu^0_{{\rm n}, \lambda}\textbf{I}_{\rm n}\right), \,\textbf{P}_{\rm Fn}(t) \right]
    \label{rt neo nuc eom}
\end{equation}
\begin{equation}
        i\hbar\frac{\partial}{\partial t}\textbf{P}_{\rm e}(t) = \left[\textbf{F}^{\rm{NEO}}_{\rm e}(t), \,\textbf{P}_{\rm e}(t) \right]
        \label{rt neo elec eom}
\end{equation}
Cavity mode and molecular operators are multiplied using the Kronecker product $\otimes$, which expands operators in the individual molecular and cavity mode Hilbert spaces into the joint molecule-mode Hilbert space. $\textbf{I}_{\rm F}$ and $\textbf{H}_{\rm F}$ are the matrix representations of the identity operator $\hat{I}_{\rm F}$ in the mode Hilbert space and the sum of harmonic oscillator Hamiltonians $\hat{H}_{\rm F}$ given by 
\begin{equation}
    \hat{H}_{\rm F} = 
    \sum_{k,\lambda}\left[\frac{1}{2}\hat{p}_{k,\lambda}^2 + \frac{1}{2}\omega_{k,\lambda}^2\hat{q}_{k,\lambda}^2\right]
    \label{free field hamiltonian}
\end{equation}
In Eq. \ref{rt neo nuc eom}, $\textbf{q}_{k,\lambda}$ is the matrix representation of $\hat{q}_{k,\lambda}$ defined above, and $\textbf{I}_{\rm n}$ is the matrix representations of the identity operator $\hat{I}_{\rm n}$ in the molecular Hilbert space . The light--matter coupling $\varepsilon_{k,\lambda}$ between the molecular system and the mode $\{k, \lambda\}$ is
\begin{equation}
    \varepsilon_{k,\lambda} = \frac{\omega_{k,\lambda}}{\sqrt{\Omega\epsilon_0}}
    \label{coupling definition}
\end{equation}
The matrix $\boldsymbol{\mu}_{{\rm n}, \lambda}$ is the matrix representation of the single-particle quantum nuclear dipole moment operator's component in the polarization direction $\lambda$. We subtract the component of the permanent quantum nuclear dipole moment in the cavity mode direction, $\mu^0_{{\rm n}, \lambda}$, from the dipole moment operator to ensure that the mode is not perturbed by the permanent dipole moment at $t = 0$.

The electronic and nuclear time-dependent Kohn-Sham matrices, $\textbf{F}^{\rm{NEO}}_{\rm e}(t)$ and $\textbf{F}^{\rm{NEO}}_{\rm n}(t)$, are given by
\begin{equation}
    \textbf{F}^{\rm{NEO}}_{\rm e}(t) = \textbf{H}^{\rm e}_{\rm core} + \textbf{J}^{\rm ee}[\textbf{P}_{\rm e}(t)] + \textbf{V}_{\rm xc}^{\rm ee}[\textbf{P}_{\rm e}(t)] - \textbf{J}^{\rm en}[\textbf{P}_{\rm n}(t)] -  \textbf{V}_{\rm c}^{\rm en}[\textbf{P}_{\rm e}(t),\textbf{P}_{\rm n}(t)] 
    \label{rt neo elec KS}
\end{equation}
\begin{equation}
    \textbf{F}^{\rm{NEO}}_{\rm n}(t) = \textbf{H}^{\rm n}_{\rm core}+ \textbf{J}^{\rm nn}[\textbf{P}_{\rm n}(t)] + \textbf{V}^{\rm nn}_{\rm xc}[\textbf{P}_{\rm n}(t)] - \textbf{J}^{\rm ne}[\textbf{P}_{\rm e}(t)] -  \textbf{V}_{\rm c}^{\rm ne}[\textbf{P}_{\rm n}(t),\textbf{P}_{\rm e}(t)] 
    \label{rt neo nuc KS}
\end{equation}
In Eq. \ref{rt neo nuc KS}, $\textbf{H}^{\rm n}_{\rm core}$ is the core Hamiltonian describing the quantum nuclear kinetic energy and the Coulomb interaction between quantum and classical nuclei; $\textbf{J}^{\rm nn}[\textbf{P}_{\rm n}(t)]$ describes the Coulomb interaction between quantum nuclei; $\textbf{V}^{\rm nn}_{\rm xc}[\textbf{P}_{\rm n}(t)]$ describes the exchange-correlation interaction between quantum nuclei; $\textbf{J}^{\rm ne}[\textbf{P}_{\rm e}(t)]$ describes the Coulomb interaction between quantum nuclei and electrons; and $\textbf{V}_{\rm c}^{\rm ne}[\textbf{P}_{\rm n}(t),\textbf{P}_{\rm e}(t)]$ describes the correlation energy between quantum nuclei and electrons. The electronic quantities in Eq. \ref{rt neo elec KS} are defined analogously.

Note that the cavity mode only couples directly to the quantum nuclear dipole moment and does not couple directly to the electronic dipole moment. This approximation is justified if the electronic and vibrational transitions correspond to significantly different energy scales, and the cavity mode energy corresponds to vibrational excitations. Thus, this approximation is reasonable for VSC on the ground-state electronic surface, where infrared radiation probes vibrational polaritons. 
The cavity modes are coupled indirectly to the electronic degrees of freedom through the dependence of the electronic and nuclear time-dependent Kohn-Sham matrices on both the electronic and quantum nuclear density matrices, $\textbf{P}_{\rm e}(t)$ and $\textbf{P}_{\rm n}(t)$.

To compute observables for either the cavity mode or quantum nuclear subsystems, we can take the partial trace of the joint density matrix $\textbf{P}_{\rm Fn}(t)$ over the degrees of freedom of the other subsystem to obtain
\begin{equation}
    \textbf{P}_{\rm n}(t) = \textrm{Tr}_{\rm F}[\textbf{P}_{\rm Fn}(t)]
    \label{field partial trace}
\end{equation}
\begin{equation}
    \textbf{P}_{\rm F}(t) = \textrm{Tr}_{\rm n}[\textbf{P}_{\rm Fn}(t)]
    \label{matter partial trace}
\end{equation}
To measure the light--matter entanglement, we compute the von Neumann entropy $S(t)$, given by
\begin{equation}
    S(t) = -\textrm{Tr}[\textbf{P}_{\rm n}(t)\ln\textbf{P}_{\rm n}(t)] = -\textrm{Tr}[\textbf{P}_{\rm F}(t)\ln\textbf{P}_{\rm F}(t)]
    \label{entropy}
\end{equation}
The second equality is proved in the Supporting Information of Ref. \citenum{welman_light-matter_2025} and has been confirmed numerically for our simulations. In our code, we compute $S(t)$ using $\textbf{P}_{\rm n}(t)$. $S(t)$ has a minimum value of 0, corresponding to the case where $\textbf{P}_{\rm Fn}(t)$ is separable into cavity mode and quantum nuclear density matrices. The von Neumann entropy quantifies how close $\textbf{P}_{\rm Fn}(t)$ is to being separable. A larger value of $S(t)$ means that $\textbf{P}_{\rm Fn}(t)$ is further from being separable. The maximum possible value of $S(t)$ in a simulation is $S_{\max} = \ln(N_{\min})$, where $N_{\min}$ is the smaller of the cavity mode and molecular basis set sizes. Other relevant properties of $S(t)$ are discussed in Ref. \citenum{welman_light-matter_2025}. 

To obtain the mfq-RT-NEO equations of motion, we assume that the joint density matrix $\textbf{P}_{\rm Fn}(t)$ is separable at all times $t \geq 0$:
\begin{equation}
    \textbf{P}_{\rm Fn}(t) = \textbf{P}_{\rm F}(t) \otimes \textbf{P}_{\rm n}(t)
    \label{separability ansatz}
\end{equation}
This is equivalent to assuming that there is zero light--matter entanglement. Eq. 4 then separates into two coupled equations of motion, one for the cavity mode density matrix $\textbf{P}_{\rm F}(t)$ and the other for the quantum nuclear density matrix $\textbf{P}_{\rm n}(t)$. The equation of motion for the electronic density matrix $\textbf{P}_{\rm e}(t)$ given by Eq. 5 remains unchanged. The resulting equations of motion are 
\begin{equation}
    i\hbar \frac{\partial}{\partial t}\textbf{P}_{\rm F}(t) = \left[\textbf{H}_{\rm F} + \sum_{k,\lambda}\varepsilon_{k,\lambda}\textbf{q}_{k,\lambda}\mu_{{\rm n}, \lambda}(t), \:\textbf{P}_{\rm F}(t)\right]
    \label{mf rt neo field eom}
\end{equation}
\begin{equation}
    i\hbar \frac{\partial}{\partial t}\textbf{P}_{\rm n}(t) = \left[\textbf{F}^{\rm NEO}_{\rm n}(t) + \sum_{k,\lambda}\varepsilon_{k,\lambda}q_{k,\lambda}(t)\boldsymbol{\mu}_{{\rm n}, \lambda}, \:\textbf{P}_{\rm n}(t)\right] 
    \label{mf rt neo nuc eom}
\end{equation}
\begin{equation}
    i\hbar\frac{\partial}{\partial t}\textbf{P}_{\rm e}(t) = \left[\textbf{F}^{\rm NEO}_{\rm e}(t), \,\textbf{P}_{\rm e}(t) \right]
    \label{mf rt neo elec eom}
\end{equation}
Here, $q_{k,\lambda}(t) \equiv \textrm{Tr}[\textbf{P}_{\rm F}(t)\textbf{q}_{k,\lambda}]$ denotes the expectation value of $\hat{q}_{k,\lambda}$ at time $t$, and $\mu_{{\rm n}, \lambda}(t) \equiv \textrm{Tr}[\textbf{P}_{\rm n}(t)\boldsymbol{\mu}_{{\rm n}, \lambda}] - \textrm{Tr}[\textbf{P}_{\rm n}(0)\boldsymbol{\mu}_{{\rm n}, \lambda}]$ denotes the expectation value of the time-dependent quantum nuclear dipole moment operator at time $t$ minus the permanent dipole moment.

\section{Simulation Details}
The fq-RT-NEO and mfq-RT-NEO methods are implemented in a developer version of QChem \cite{epifanovsky_software_2021}. In the fq-RT-NEO method, all quantum mechanical degrees of freedom (mode, quantum nuclei, and electrons) are propagated with a modified-midpoint unitary transform (MMUT) time-propagation scheme \cite{li_time-dependent_2005} with an additional predictor-corrector scheme \cite{de_santis_pyberthart_2020} used to control numerical error during the real-time dynamics. The quantum molecular degrees of freedom in the mfq-RT-NEO method are propagated with the same scheme, while the quantum mode degrees of freedom are propagated with the time-evolution operator $\exp(-i\textbf{H}t/\hbar)$, where $\textbf{H} = \textbf{H}_{\rm F} + \sum_{k,\lambda}\varepsilon_{k,\lambda}\textbf{q}_{k,\lambda}\mu_{{\rm n}, \lambda}(t)$ with all terms defined above. The classical nuclei are fixed at specified geometries in both methods.   

We simulated HCN under VSC with the electrons and the proton treated quantum mechanically. The calculations were performed with the cc-pVDZ \cite{dunning_gaussian_1989} electronic basis set and  the even-tempered 8s8p8d \cite{yang_development_2017} protonic basis set with exponents ranging from $2\sqrt{2}$ to 32. The B3LYP \cite{lee_development_1988, becke_new_1998, becke_density-functional_1988} electronic exchange-correlation functional and the epc17-2 \cite{yang_development_2017, brorsen_multicomponent_2017} electron-proton correlation functional were used. The  C-N bond length was taken to be 1.16 \r{A}, and the distance between the proton basis function center and the carbon nucleus was taken to be 1.07 \r{A}.
The cavity mode was described with four number basis functions $\ket{i}, i = 0, 1,2,3$, which is the largest computationally tractable basis set that can be used with the aforementioned molecular basis sets given our resources.  

At $t = 0$ for the fq-RT-NEO method, we initialized the joint cavity mode--nuclear density matrix $\textbf{P}_{\rm Fn}(0)$ as a separable product of cavity mode and quantum nuclear density matrices: $\textbf{P}_{\rm Fn}(0) = \textbf{P}_{\rm F}(0) \otimes \textbf{P}_{\rm n}(0)$. We chose $\textbf{P}_{\rm F}(0) = \ket{1}\bra{1}$ (i.e., we initialized the cavity mode in the first harmonic oscillator excited state populated by a single photon), and we chose $\textbf{P}_{\rm n}(0)$ to be the self-consistent field (SCF) ground state quantum nuclear density matrix from a NEO-DFT calculation. We also chose $\textbf{P}_{\rm e}(0)$ to be the SCF ground state electronic density matrix from the same NEO-DFT calculation. Identical initial conditions were used for the mfq-RT-NEO calculation, with the only difference being that we did not need to initialize a joint mode--nuclear density matrix because this method propagates those subsystems separately. We propagated the real-time dynamics with a time step of $\Delta t = 0.04$ a.u. and light--matter coupling of $8 \times 10^{-4}$ a.u. for both methods. All power spectra $P\{f(t)\}$ of real-time data $f(t)$ were computed as $P\{f(t)\} = |\mathcal{F}\{f(t)e^{-\gamma t}\}|$, where  $\mathcal{F}\{...\}$ denotes the Pad\'{e} approximation\cite{de_santis_pyberthart_2020, goings_realtime_2018} to the Fourier transform. We used a small damping of $\gamma = 10^{-5}$ a.u., giving a linewidth of $1.7 \times10^{-3}$ eV = 13.8 cm\textsuperscript{-1} to all peaks. 
\section{Results}
\begin{figure}
\centering
\includegraphics[scale=0.9, trim=2 4 2 2, clip]{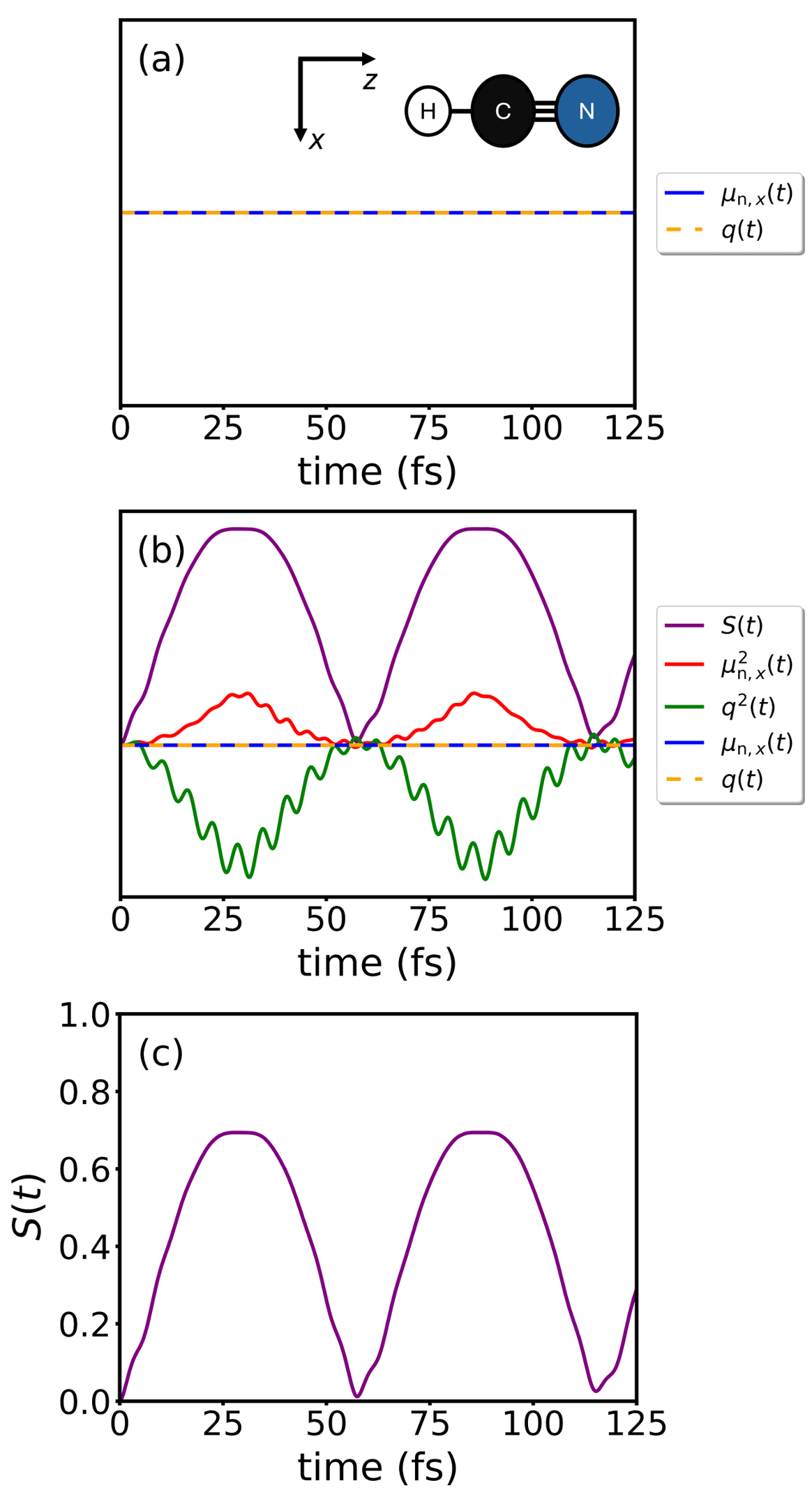}
    \caption{Comparison of time-dependent properties computed with mfq- and fq-RT-NEO dynamics applied to HCN. (a) mfq-RT-NEO dynamics of $\mu_{{\rm n}, x}(t)$ and $q(t)$. No significant oscillations are observed. No units are given on the $y$-axis because these observables have different units. (b) fq-RT-NEO dynamics of $\mu_{{\rm n}, x}(t)$ and $q(t)$, as well as $S(t)$, $\mu^2_{{\rm n}, x}(t)$ and $q^2(t)$.  No units are given on the $y$-axis because these observables have different units. (c) fq-RT-NEO dynamics of $S(t)$. The maximum computed value of $S(t)$ is $\sim$ 0.67.}
    \label{data}
\end{figure}

We consider the HCN molecule shown in Figure \ref{data}a. At $t = 0$, we initialize the molecule in its NEO-DFT ground state and populate the cavity mode with one photon, corresponding to initializing it in the first harmonic oscillator excited state or single-photon Fock state. The molecule is coupled to an $x$-polarized lossless cavity mode tuned to be in resonance with the $x$-direction bending mode of the quantum proton, which has frequency $\omega_{\rm c} = 2803 \textrm{ cm}^{-1}$ at this level of theory, and the light--matter coupling strength is $\varepsilon = 8 \times 10^{-4}$ a.u.  We then propagate the real-time excited state dynamics with either the mfq-RT-NEO or fq-RT-NEO method.

First we consider the results from the mfq-RT-NEO dynamics simulation. Figure \ref{data}a shows the time evolution of the cavity mode coordinate, $q(t) \equiv \textrm{Tr}[\textbf{P}_{\rm F}(t)\textbf{q}_{k,x}] - \textrm{Tr}[\textbf{P}_{\rm F}(0)\textbf{q}_{k,x}]$, and quantum nuclear dipole moment, $\mu_{{\rm n}, x}(t) \equiv \textrm{Tr}(\textbf{P}_{\rm n}(t)\boldsymbol{\mu}_{{\rm n}, x}) - \textrm{Tr}(\textbf{P}_{\rm n}(0)\boldsymbol{\mu}_{{\rm n}, x})$, where $k$ is the magnitude of the cavity mode wavevector as defined in the context of Eq. \ref{qed hamiltonian}, and the cavity mode is polarized in the $x$-direction. Both quantities remain constant for the duration of the dynamics, suggesting that at the mfq-RT-NEO level of theory, energy exchange does not occur between the cavity mode and molecular subsystems when the mode subsystem is initialized in a Fock state and the molecule is initialized in its ground state. In other words, a molecular polariton is not formed because there is no feedback interaction between the two subsystems. This result can be easily rationalized by considering the mfq-RT-NEO equations given by Eqs. 14 and 15. At $t = 0$, $\mu_{{\rm n}, x}(t) = 0$, so according to Eq. 14, the field density matrix $\textbf{P}_{\rm F}(t)$ is unchanged. By the same reasoning, given that $q(t) = 0$ at $t = 0$, the quantum nuclear density matrix $\textbf{P}_{\rm n}(t)$ is also unchanged. Without a further perturbation, the density matrices are preserved, leaving all observables unchanged from their initial values. We note that a similar argument in the context of a model polaritonic system with a classical cavity mode was recently made in Ref. \citenum{simko_twin_2025}.

We now examine whether the absence of coherent molecule--mode energy transfer and polariton formation persists under fq-RT-NEO dynamics. Figure \ref{data}b shows the time evolution of $q(t)$, $\mu_{{\rm n}, x}(t)$, and the von Neumann entropy $S(t)$. Once again, $q(t)$ and $\mu_{{\rm n}, x}(t)$ remain essentially unchanged from their initial values. The persistence of this result is surprising given that neither $\hat{q}_{k, \lambda}$ nor $\hat{\mu}_{{\rm n}, x}$ commute with $\hat{H}_{\rm QED}$, and hence their expectation values are not  expected to be constants of motion. These results will be considered in more detail in the Discussion. 

In contrast, the von Neumann entropy $S(t)$ shows clear oscillations on a timescale of $\sim $59 fs in the fq-RT-NEO dynamics. These oscillations are of significant amplitude. The maximum possible value of the von Neumann entropy that could be theoretically attained in this simulation is given by the natural logarithm of the smaller of the quantum nuclear and cavity mode basis sets. The cavity mode basis set is smaller with only four functions, so $S_{\rm \max} = \ln(4) = 1.38$. Figure 1c shows that the maximum value of $S(t)$ computed in the fq-RT-NEO calculation is $\sim 0.67$, or approximately 48\% of $S_{\rm max}$. Although not a perfect comparison, it is still worth noting that this degree of light--matter entanglement is far larger than the degree of light--matter entanglement observed when the cavity mode was initialized in a coherent state with a small initial displacement in the linear response regime \cite{welman_light-matter_2025}. This result suggests that coherent energy transfer between the molecule and cavity mode subsystems may be occurring, even if the cavity mode coordinate and quantum nuclear dipole moment are constant throughout the dynamics. We hypothesized that this transfer process might manifest in oscillations of the expectation values of the operators $\hat{q}^2_{k,x}$ and $\hat{\mu}^2_{{\rm n}, x}$. To interrogate this possibility, we also plot the dynamics of $q^2(t) \equiv \textrm{Tr}[\textbf{P}_{\rm F}(t)\textbf{q}^2_{k,x}] - \textrm{Tr}[\textbf{P}_{\rm F}(0)\textbf{q}^2_{k,x}]$ and $\mu^2_{{\rm n}, x}(t) \equiv \textrm{Tr}[\textbf{P}_{\rm n}(t)\boldsymbol{\mu}^2_{{\rm n}, x}] - \textrm{Tr}[\textbf{P}_{\rm n}(0)\boldsymbol{\mu}^2_{{\rm n}, x}]$ in Figure \ref{data}b. Both quantities display slow oscillations, again on a timescale of $\sim $ 59 fs, as well as fast oscillations on a timescale of $\sim$ 6 fs. The fast oscillation timescale is almost exactly half of that corresponding to the oscillation frequency $\omega_{\rm c}$ of the cavity mode, which is $2\pi/\omega_{\rm c} \approx 12 $ fs. 
%We note that this comparison of $q(t)$ and $q^2(t)$ parallels recent analysis in Ref. \citenum{simko_twin_2025}. 

\begin{figure}[H]
\centering
\includegraphics[scale=0.9, trim=2 2 2 2, clip]{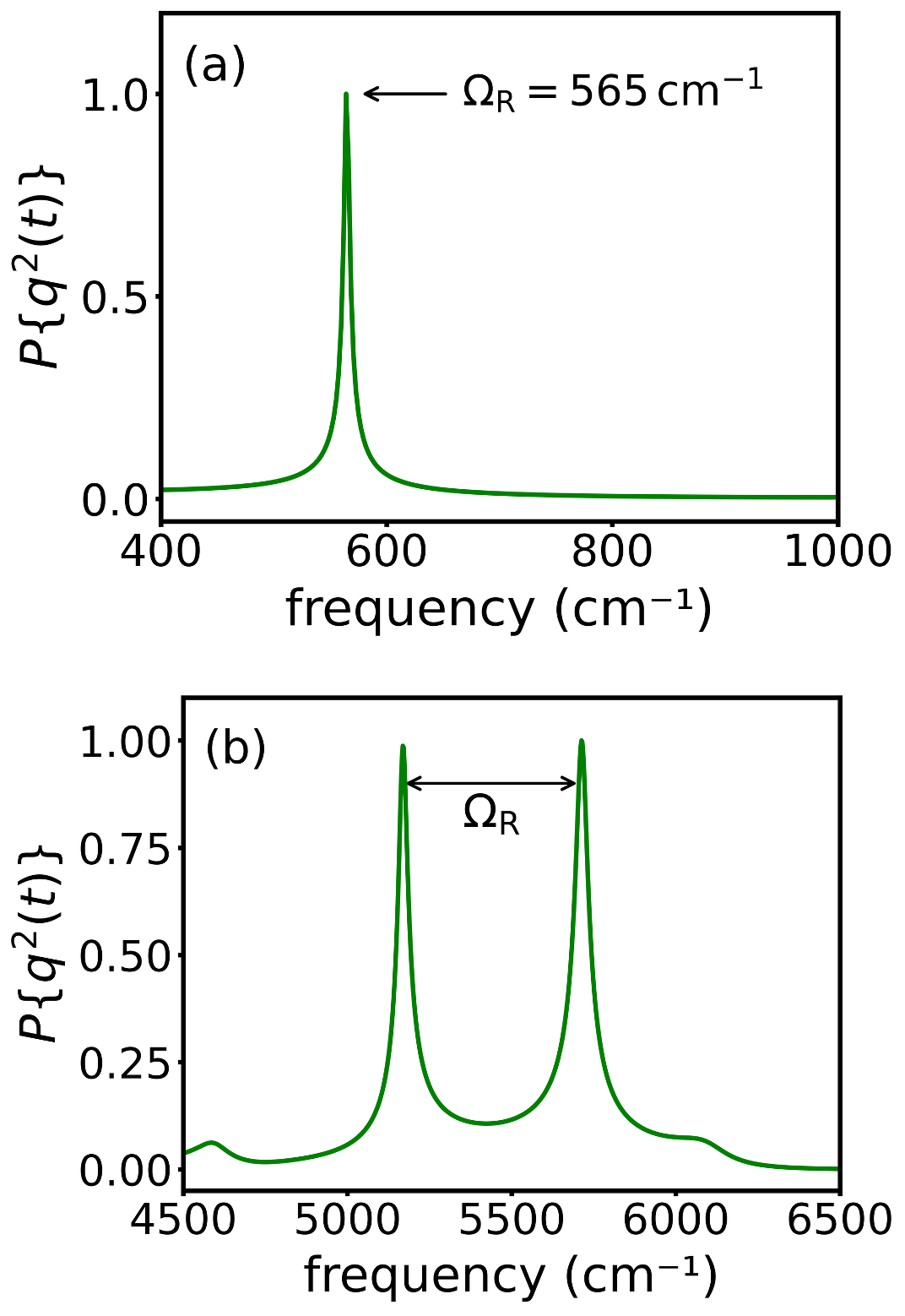}
    \caption{Power spectrum of $q^2(t)$ obtained with fq-RT-NEO dynamics applied to HCN for the (a) 400 - 1000 cm\textsuperscript{-1} region and (b) 4500 - 6500 cm\textsuperscript{-1} region. In both spectra, the largest signal amplitude has been scaled to a value of 1.}
    \label{q square FT}
\end{figure}

In order to quantify the frequencies more precisely, we considered the power spectrum of $q^2(t)$. Figure \ref{q square FT}a shows the power spectrum of $q^2(t)$ in the spectral region of 400 to 1000 cm$^{-1}$, exhibiting a single peak at 565 cm$^{-1}$. This frequency corresponds to an oscillation timescale of 59 fs, matching the slow frequency observed in the oscillations of $S(t)$, $q^2(t)$, and $\mu^2_{{\rm n}, x}(t)$. We will interpret this frequency as the Rabi frequency and denote it as $\Omega_{\rm R}$. Figure \ref{q square FT}b shows the power spectrum of $q^2(t)$ in the spectral region of 4500 to 6500 cm$^{-1}$. The power spectrum in this region shows a pair of peaks centered at 5437 cm$^{-1}$, which agrees with twice the cavity mode frequency $2\omega_{\rm c}$ to within 3\%. This frequency corresponds to a timescale of 6 fs, matching the timescale of the fast oscillations observed in the dynamics of $q^2(t)$ and $\mu^2_{{\rm n}, x}(t)$. The splitting between the two peaks is 545 cm$^{-1}$, which agrees with $\Omega_{\rm R}$ to within 4\%. Although it is reassuring that $\Omega_{\rm R}$ appears in the spectrum as a splitting between a pair of peaks, we would usually expect to see that pair of peaks centered around $\omega_{\rm c}$ and not $2\omega_{\rm c}$. Interestingly, there are no features in the power spectrum of $q^2(t)$ in the region of the cavity mode frequency $\omega_{\rm c}$.  These observations will be analyzed in more detail in the Discussion.

\begin{figure}[H]
\centering
\includegraphics[scale=1.0, trim=2 2 2 2, clip]{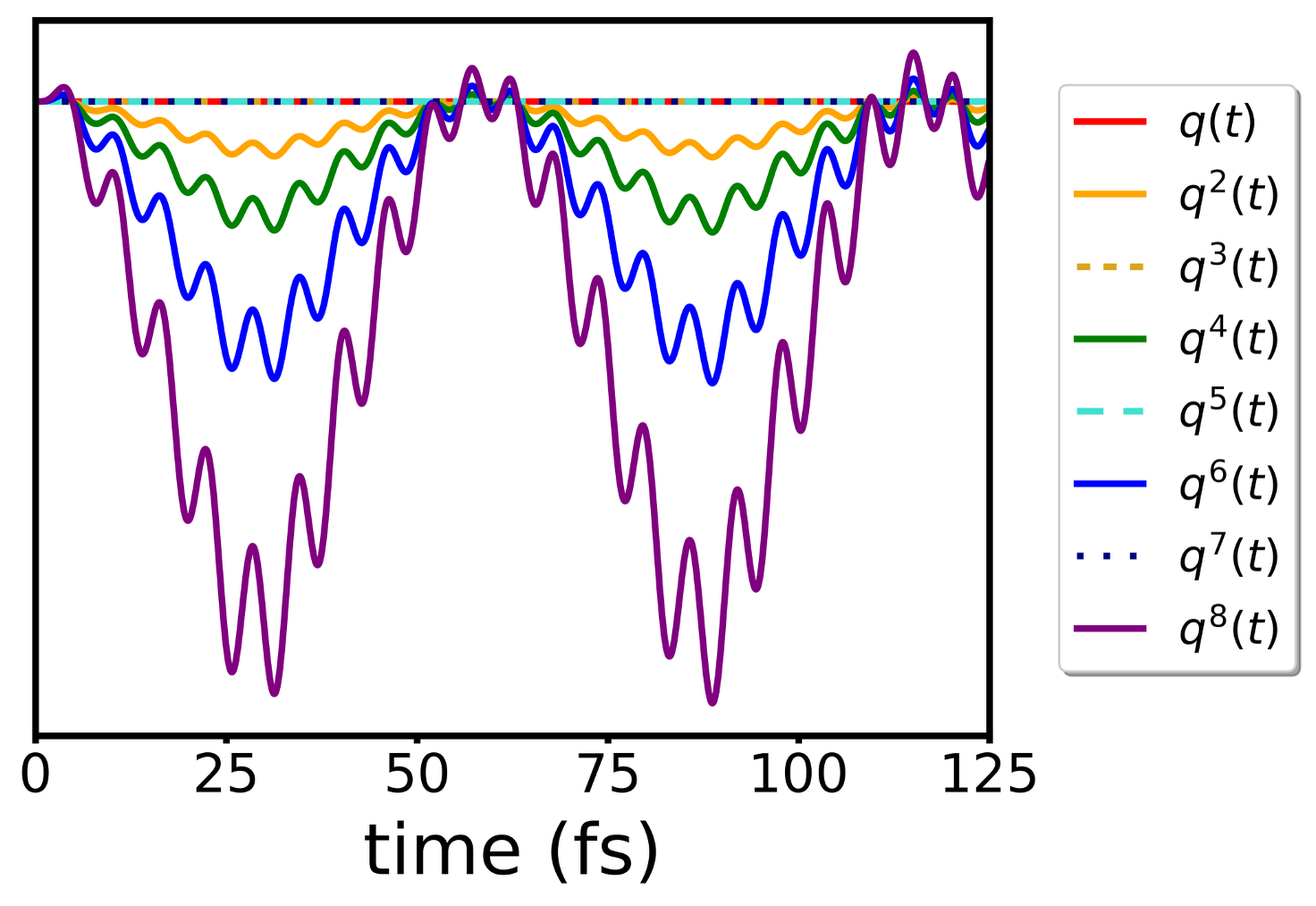}
    \caption{fq-RT-NEO dynamics of $q^n(t)$ for $n = 1, ..., 8$. No units are given on the $y$-axis because these observables have different units.}
    \label{q power plot}
\end{figure}

The observation that the first-order quantities $q(t)$ and $\mu_{{\rm n}, x}(t)$ are constant, while the quadratic quantities $q^2(t)$ and $\mu^2_{{\rm n}, x}(t)$  oscillate, suggests the possibility of a more general trend,  whereby the expectation values of odd powers of $\hat{q}_{k,x}$ and $\hat{\mu}_{{\rm n}, x}$ are constant and the expectation values of even powers of these operators oscillate. To test this hypothesis, we plot $q^n(t) \equiv \textrm{Tr}[\textbf{P}_{\rm F}(t)\textbf{q}^n_{k,x}] - \textrm{Tr}[\textbf{P}_{\rm F}(0)\textbf{q}^n_{k,x}]$ for $n = 1 ... 8$ in Figure \ref{q power plot} and find that this trend does hold. The expectation values of even powers oscillate, while the expectation values of odd powers of $\hat{q}_{k,\lambda}$ remain constant. In Figure S1 in the Supporting Information, we show the corresponding plot for $\mu^n_{{\rm n}, x}(t) \equiv \textrm{Tr}[\textbf{P}_{\rm n}(t)\boldsymbol{\mu}^n_{{\rm n}, x}] - \textrm{Tr}[\textbf{P}_{\rm n}(0)\boldsymbol{\mu}^n_{{\rm n}, x}]$ for $n = 1 ... 8$. The same trend for expectation values of even and odd operator powers holds for $\hat{\mu}_{{\rm n}, x}$ as well. 
%Unlike the oscillations of $q^n(t)$, however, the amplitudes of the oscillations of  $\mu^n_{{\rm n}, \lambda}(t)$ decrease with increasing $n$. We can rationalize this trend by noting that for a charge distribution associated with a single quantized particle, expectation values of powers $n > 2$ of the dipole operator correspond to higher-order terms in a multipole expansion. The decay of oscillation amplitude with increasing $n$ simply reflects the decreasing importance of higher-order terms in the multipole expansion. 
In Figure S2, we show the power spectrum of $\mu^2_{{\rm n}, x}(t)$, which agrees qualitatively with the power spectrum of $q^2(t)$ shown in Figure \ref{q square FT}. Figure S2a exhibits a peak at 565 cm\textsuperscript{-1}, perfectly matching the $\Omega_{\rm R}$ peak in Figure \ref{q square FT}a. Similar to Figure \ref{q square FT}b, Figure S2b shows a pair of peaks separated by 555 cm\textsuperscript{-1}, which agrees with the Rabi splitting $\Omega_{\rm R} = 565$ cm\textsuperscript{-1} to within 2\%. This pair of peaks is centered at a frequency of 4918 cm\textsuperscript{-1}, which is redshifted from the center of the pair of peaks in the power spectrum of $q^2(t)$ by 519 cm\textsuperscript{-1} but still agrees with $2\omega_c = 5606$ cm\textsuperscript{-1} to within $\sim$ 12\%. Finally, we note that by the argument given previously, all of these observables would be constant at the mfq-RT-NEO level of theory.

Our mfq-RT-NEO and fq-RT-NEO dynamics initialized with the molecule in its ground state and the cavity mode in a one-photon Fock state yielded several interesting results. In the mfq-RT-NEO simulation, coherent energy transfer between the molecule and the cavity mode was entirely suppressed, and polariton formation did not occur. Coherent energy transfer and significant light--matter entanglement did occur in the fq-RT-NEO simulation, but this interaction manifested only in the oscillations of even powers of the cavity mode coordinate and quantum nuclear dipole operators. Furthermore, the power spectra of these nonzero expectation values displayed  spectral features that did not include the splitting of a bare molecular peak at the cavity mode frequency  into two polariton peaks. Below, we will consider all of these results in more detail, aiming to facilitate our understanding through comparisons with analytically tractable quantum optics models.

\newpage 

\section{Discussion}

In this section, we will employ quantum optical models to understand the key observations presented above in the fq-RT-NEO results: the lack of oscillations in $q(t)$ and $\mu_{{\rm n}, x}(t)$, the spectral features of the power spectrum of $q^2(t)$ shown in Figure \ref{q square FT}, and the trend in expectation values of $q^n(t)$ and $\mu^n_{{\rm n}, x}(t)$ for even and odd powers of $n$ shown in Figure \ref{q power plot} and Figure S1. First, we will use a quantum Rabi model Hamiltonian in a minimal basis to better understand the lack of oscillations in the dynamics of $q(t)$ and $\mu_{{\rm n}, x}(t)$ and determine the criteria needed to recover the canonical feature of cavity-modified spectra, namely the splitting of a bare molecular transition into two polariton peaks. We will then consider the same model in a larger basis in order to explain the trend in expectation values of $q^n(t)$ and $\mu^n_{{\rm n}, x}(t)$ for even and odd powers of $n$, as well as the spectral features and oscillation frequencies in Figures \ref{q square FT} and \ref{q power plot} and Figure S1. 

Throughout this section, we will make extensive use of dressed-state kets in the joint light--matter Hilbert space denoted as $\ket{FM}$. We will adopt the convention that the quantum number $F$ determines the number of photons populating the cavity mode, while $M$ corresponds to the $M^{\rm th}$ eigenstate of the bare molecular subsystem.

\subsection{Criteria for Oscillations of $q(t)$ and $\mu_{{\rm n}, x}(t)$}

We will first examine the requirements for oscillations of the expectation values of the cavity mode coordinate and molecular dipole moment operators. This understanding will clarify the criteria for the initial conditions required to observe the bare molecular peak split into two polariton peaks. 
%For simplicity, we first summarize our argument here as follows. The minimal basis QED Hamiltonian we will consider below has the property that it is block diagonal in the dressed-state basis $\ket{FM}$, while the cavity mode coordinate and molecular dipole moment operators have only off-diagonal blocks when written in the same basis. Thus, in order to observe  oscillations of the cavity mode coordinate and molecular dipole moment expectation values, we must initialize the system in a superposition of states from each block. The state $\ket{10}$, which corresponds to our fq-RT-NEO initial condition, is a state in only one of the blocks, so we will not observe any oscillations of either expectation value. In contrast, if we were to initialize the system in a superposition of states from the two blocks of the model QED Hamiltonian, then we would observe oscillating expectation values with frequencies given by the energy difference between the ground state associated with the QED Hamiltonian  and the polariton states. These frequencies will manifest in a power spectrum of those expectation values as a single bare molecular peak splitting into two polariton peaks. This is analogous to the phenomenology observed with a coherent state initial condition for a quantum cavity mode \cite{welman_light-matter_2025}. 

We consider a model quantum Rabi Hamiltonian of the form
\begin{equation}
\begin{split}
    \hat{H} &= \hat{H}_{\rm F} + \hat{H}_{\rm M} + \varepsilon\hat{q}\hat{\mu}\\
    &= \hat{H}_{\rm F} + \hat{H}_{\rm M} + \varepsilon q_0(\hat{a}^\dagger + \hat{a})\mu_0(\hat{b}^\dagger + \hat{b}) \\
    &= \omega\hat{a}^\dagger\hat{a} + \omega\hat{b}^\dagger\hat{b} + g(\hat{a}^\dagger\hat{b}^\dagger + \hat{a}^\dagger\hat{b} + \hat{a}\hat{b}^\dagger + \hat{a}\hat{b})
\end{split}
\label{two level quantum rabi}
\end{equation}
Here, we have taken $\hbar = 1$ in atomic units. As before, the subscript ``F" denotes field, which refers to the cavity mode. We define $\hat{a}$ ($\hat{a}^\dagger$) as the photon annihilation (creation) operator and $\hat{b}$ ($\hat{b}^\dagger$) as the molecular excitation annihilation (creation) operator. In this notation, $\hat{q} = q_0(\hat{a}^\dagger + \hat{a})$ is the mode coordinate operator, and $\hat{\mu} = \mu_0(\hat{b}^\dagger + \hat{b})$ is the molecular dipole operator, where $q_0$ and $\mu_0$ are constants that  give their respective operators the appropriate dimensions. The light--matter coupling is given by $g = \varepsilon q_0\mu_0$, where $\varepsilon$ is a constant with appropriate dimensions, comparable to $\varepsilon_{k, \lambda}$ in Eq. \ref{coupling definition}, such that $g$ has dimensions of energy. $\hat{H}_{\rm F} = \omega\hat{a}^\dagger\hat{a}$ and $\hat{H}_{\rm M} = \omega\hat{b}^\dagger\hat{b}$  are the bare cavity mode and molecular Hamiltonians, respectively. For simplicity, we assume the cavity mode to be in resonance with the molecular transition at energy $\omega$ in atomic units. Note that in this model, the vibrational mode is assumed to be harmonic. Finally, we point out that even though we take $\hat{b}$ and $\hat{b}^\dagger$ to be bosonic operators, our subsequent results would remain valid if these bosonic operators were replaced by fermionic creation and annihilation operators.

In this section, we consider this Hamiltonian in the basis of four dressed states: $\ket{00}, \ket{11}, \ket{01}, $ and $\ket{10}$. In this basis, \textbf{H}, the matrix representation of $\hat{H}$, factors into two $2\times2$ blocks:
\begin{equation}
    \textbf{H} = 
    \begin{pmatrix}
        0 & g & & \\
        g & 2\omega & & \\
         & & \omega & g \\
         & & g & \omega \\
    \end{pmatrix}
    = 
    \begin{pmatrix}
        \textbf{H}_{\rm CR} & \\
         & \textbf{H}_{\rm JC} \\
    \end{pmatrix}
    \label{two level quantum rabi matrix}
\end{equation}
The lower right block, $\textbf{H}_{\rm JC}$, is the Hamiltonian of the Jaynes-Cummings model. It can be understood as the contribution of the terms $\hat{a}^\dagger\hat{b}$ and $\hat{a}\hat{b}^\dagger$ in Eq. \ref{two level quantum rabi}. The upper left block, $\textbf{H}_{\rm CR}$, corresponds to the contribution of the double excitation and de-excitation counter-rotating terms $\hat{a}^\dagger\hat{b}^\dagger$ and $\hat{a}\hat{b}$ in Eq. \ref{two level quantum rabi}. Application of the rotating wave approximation (RWA) to Eq. \ref{two level quantum rabi} would eliminate the counter-rotating terms and by extension the off-diagonal elements of $\textbf{H}_{\rm CR}$. We also note that for the four dressed states $\ket{FM}$, $F-M$ is even for states in the counterrotating block and is odd for states in the Jaynes-Cummings block. We will expand on this observation in more detail below. 

We now diagonalize both $\textbf{H}_{\rm JC}$ and $\textbf{H}_{\rm CR}$. The differences between the eigenenergies will determine the possible frequencies at which an observable evolving under the Hamiltonian $\textbf{H}$ could oscillate. The eigenenergies and eigenstates for $\textbf{H}_{\rm JC}$ are
\begin{equation}
    E^{\rm JC}_{\pm} = \omega \pm g
\end{equation}
\begin{equation}
    \ket{\psi_{\pm}^{\rm JC}} = \frac{1}{\sqrt{2}}\left(\ket{01} \pm \ket{10}\right)
\end{equation}
The Rabi splitting is given by $\Omega_{\rm R} = E^{\rm JC}_{+} - \textrm{E}^{\rm JC}_{-} = 2g$. The eigenenergies and eigenstates for $\textbf{H}_{\rm CR}$ are
\begin{equation}
    E^{\rm CR}_{\pm} = \omega \pm \sqrt{\omega^2 + g^2}
\end{equation}
\begin{equation}
    \ket{\psi_{\pm}^{\rm CR}} = -\frac{E^{\rm CR}_{\mp}}{gN_{\pm}}\ket{00} + \frac{1}{N_{\pm}}\ket{11}
\end{equation}
where we introduce the normalization constants $N_{\pm} \equiv \sqrt{1+(E^{\rm CR}_{\mp}/g)^2}$. Note that $\ket{\psi_{-}^{\rm CR}}$ is the ground state in this basis. If we introduce transformation matrices
\begin{equation}
    \textbf{U}_{\rm JC} = 
    \begin{pmatrix}
        \frac{1}{\sqrt{2}} & \frac{1}{\sqrt{2}} \\
        \frac{1}{\sqrt{2}} & -\frac{1}{\sqrt{2}} \\
    \end{pmatrix}
\end{equation}
\begin{equation}
    \textbf{U}_{\rm CR} = 
    \begin{pmatrix}
        -\frac{E^{\rm CR}_{-}}{gN_+} & -\frac{E^{\rm CR}_{+}}{gN_-} \\
        \frac{1}{N_+} & \frac{1}{N_-} \\
    \end{pmatrix}
\end{equation}
then we can transform any matrix from the original basis of four dressed states to the eigenbasis of $\textbf{H}$, $\ket{\psi_{\pm}^{\rm CR}}$ and $\ket{\psi_{\pm}^{\rm JC}}$ (hereafter denoted the ``energy eigenbasis"), using the total transformation matrix given by the direct sum
\begin{equation}
    \textbf{U}_{\rm T} = \textbf{U}_{\rm CR} \oplus \textbf{U}_{\rm JC} = 
    \begin{pmatrix}
        \textbf{U}_{\rm CR} & \\
         & \textbf{U}_{\rm JC} \\
    \end{pmatrix}
    \label{transformation matrix}
\end{equation}

We now transform the matrices $\textbf{q}$ and $\boldsymbol{\mu}$, the matrix representations of the operators $\hat{q}$ and $\hat{\mu}$, from the dressed-state basis to the energy eigenbasis using $\textbf{U}_{\rm T}$. In the dressed-state basis, these matrices are given by 
\begin{equation}
    \textbf{q} = q_0
    \begin{pmatrix}
         & & & 1 \\
         & & 1 &  \\
         & 1 & &  \\
         1 & & &  \\
    \end{pmatrix}
    = q_0 (\boldsymbol{\sigma}_x \otimes \boldsymbol{\sigma}_x)
    \label{coordinate dressed}
\end{equation}
\begin{equation}
    \boldsymbol{\mu} = \mu_0
    \begin{pmatrix}
         & & 1 &  \\
         & &  & 1 \\
         1 &  & &  \\
          & 1 & &  \\
    \end{pmatrix}
     = \mu_0(\boldsymbol{\sigma}_x \otimes \textbf{I}_2)
    \label{dipole dressed}
\end{equation}
where $\boldsymbol{\sigma}_x$ is the Pauli $x$-matrix and $\textbf{I}_2$ is the 2$\times$2 identity matrix.
Transforming to the energy eigenbasis, we obtain
\begin{equation}
    \textbf{q} \rightarrow \textbf{U}_{\rm T}^\dagger\textbf{q}\textbf{U}_{\rm T} = q_0
    \begin{pmatrix}
          &\textbf{U}_{\rm CR}^\dagger\boldsymbol{\sigma}_x\textbf{U}_{\rm JC}&\\
         \textbf{U}_{\rm JC}^\dagger\boldsymbol{\sigma}_x\textbf{U}_{\rm CR}& &\\
    \end{pmatrix}
    \label{coordinate energy eigenbasis}
\end{equation}

\begin{equation}
    \boldsymbol{\mu} \rightarrow \textbf{U}_{\rm T}^\dagger\boldsymbol{\mu}\textbf{U}_{\rm T} = \mu_0
    \begin{pmatrix}
          &\textbf{U}_{\rm CR}^\dagger\textbf{U}_{\rm JC}&\\
         \textbf{U}_{\rm JC}^\dagger\textbf{U}_{\rm CR}& &\\
    \end{pmatrix}
    \label{dipole energy eigenbasis}
\end{equation}

We can immediately see from the block diagonal structures of Eqs. \ref{coordinate energy eigenbasis} and \ref{dipole energy eigenbasis} why there are no oscillations of $q(t)$ and $\mu(t)$ in our fq-RT-NEO simulations when the mode is initialized in a single-photon Fock state and the molecular system is initialized in its ground state, even though neither $\hat{\mu}$ nor $\hat{q}$ commute with $\hat{H}$. In the model treatment here, the only nonzero matrix elements of $\textbf{q}$ and $\boldsymbol{\mu}$ are between the Jaynes-Cummings energy eigenstates $\ket{\psi_\pm^{\rm JC}}$ and the counterrotating energy eigenstates $\ket{\psi_\pm^{\rm CR}}$. Any state $\ket{\Psi(t)}$ of the joint light--matter system can be written as a linear combination of these four states:
\begin{equation}
\begin{split}
    \ket{\Psi(t)} = c^{\rm JC}_+\exp(-iE_+^{\rm JC}t)\ket{\psi_+^{\rm JC}} + c^{\rm JC}_-\exp(-iE_-^{\rm JC}t)\ket{\psi_-^{\rm JC}} \\
    + \, c^{\rm CR}_+\exp(-iE_+^{\rm CR}t)\ket{\psi_+^{\rm CR}} + c^{\rm CR}_-\exp(-iE_-^{\rm CR}t)\ket{\psi_-^{\rm CR}}
\end{split}
\label{general light-matter two level wavefunction}
\end{equation}
From Eq. \ref{general light-matter two level wavefunction}, we can see that any time-dependent expectation value $A(t) = \braket{\Psi(t)|\hat{A}|\Psi(t)}$ of an operator $\hat{A}$ will be a sum of terms of the form $c_ic_j^*\exp(-i(E_i - E_j)t)\braket{j|\hat{A}|i}$, where $\bra{j}$ and $\ket{i}$ are energy eigenstates. These terms will oscillate with a frequency $(E_i-E_j)$, the difference in energy between $\ket{i}$ and $\ket{j}$, assuming that $c_i$ and $c_j$, the coefficients of $\ket{i}$ and $\ket{j}$ at $t = 0$, as well as  $\braket{j|\hat{A}|i}$, are all nonzero. Based on this analysis and the structure of the matrices $\textbf{q}$ and $\boldsymbol{\mu}$ given in Eqs. \ref{coordinate energy eigenbasis} and \ref{dipole energy eigenbasis}, we conclude that we will only observe oscillations in $q(t)$ and $\mu(t)$ if the initial state of the system can be written as a superposition of Jaynes-Cummings and counterrotating energy eigenstates with nonzero contributions from both types of eigenstates. The model initial condition corresponding to the initial condition of the fq-RT-NEO calculation in this work is the dressed state $\ket{10} = \frac{1}{\sqrt{2}}\left( \ket{\psi_+^{\rm JC}} - \ket{\psi_-^{\rm JC}} \right)$. This state is exclusively a superposition of Jaynes-Cummings eigenstates, and thus we do not observe any oscillations in $q(t)$ and $\mu_{{\rm n}, x}(t)$. The oscillations of expectation values of higher powers of the operators $\hat{q}$ and $\hat{\mu}$, denoted  $q^n(t)$ and $\mu^n(t)$ with $n > 1$, will be considered in the next section. 

If the initial state of the joint light--matter system can be written as a superposition of Jaynes-Cummings and counterrotating energy eigenstates with nonzero contributions from both types, then we will observe oscillations in $q(t)$ and $\mu(t)$. As pointed out above, the frequencies of these oscillations will be given by the energy differences between the Jaynes-Cummings and counterrotating energy eigenstates. We will restrict our attention to positive energy differences without loss of generality.  We first consider the energy differences between the Jaynes-Cummings polariton states and the ground state
\begin{equation}
    E^{\rm JC}_{\pm} - E^{\rm CR}_{-} = \sqrt{\omega^2 + g^2} \pm g
\end{equation}
Assuming that $\omega >> g$, we can expand this difference in powers of the small quantity $g/\omega << 1$ to obtain a simple expression for the location of the peaks in the power spectra of $q(t)$ and $\mu(t)$:
\begin{equation}
\begin{split}
    E^{\rm JC}_{\pm} - E^{\rm CR}_{-} &= \sqrt{\omega^2 + g^2} \pm g \\
    &= \omega \sqrt{1 + \left(\frac{g}{\omega}\right)^2} \pm g \\
    &= \omega \left(1 + \frac{1}{2}\left(\frac{g}{\omega}\right)^2 + ... \right) \pm g\\
    &= \omega \pm g + O((g/\omega)^2)
\end{split}
\end{equation}
We therefore find that if our model system is initialized in a superposition of Jaynes-Cummings and counterrotating energy eigenstates with nonzero contributions from both types, then the power spectra of $q(t)$ and $\mu(t)$ exhibit a pair of peaks split by the Rabi splitting $\Omega_{\rm R} = 2g$, with the center of the peaks at the bare molecular transition energy $\omega$. Higher-order corrections to these energies are  proportional to terms that are second order in the small quantity $(g/\omega)$. This same analysis can also be carried out for the other set of positive energy differences, $E^{\rm CR}_{+} - E^{\rm JC}_{\pm}$, with identical results. 

Finally, we observe that within the framework of this model and basis set, initializing the total light--matter system in a superposition of Jaynes-Cummings and counterrotating eigenstates is equivalent to the case in which at least one of the light or matter subsystems is in a superposition of its ground and excited states. In Ref. \citenum{welman_light-matter_2025}, we initialized the joint light--matter system for our fq-RT-NEO calculations as a separable product of a coherent state of the cavity mode and the molecular ground state. A coherent state is by definition a superposition of ground and excited states, so we expect to see a bare molecular transition peak split into two peaks separated by the Rabi splitting. Since a classical mode shows qualitatively similar behavior to a quantum mode prepared in a coherent state with the analogous initial coordinate and momentum values, we expect this same phenomenon to appear with a classical mode as well. This behavior was observed for both the classical and quantum cavity mode RT-NEO simulations in Ref. \citenum{welman_light-matter_2025}. 

\subsection{Oscillations of $q^n(t)$ and $\mu^n(t)$ for even $n$}
The power rule for even and odd expectation values of $q^n(t)$ and $\mu^n(t)$ shown by the data in Figure \ref{q power plot} and Figure S1 can be understood by considering the quantum Rabi Hamiltonian given in Eq. \ref{two level quantum rabi}, but \textit{in a larger basis set}. To understand the need to use a larger basis set, we look at powers of the matrix representations $\textbf{q}$ and $\boldsymbol{\mu}$ in the four-state energy eigenbasis that we have employed thus far. Using Eqs. \ref{coordinate energy eigenbasis} and \ref{dipole energy eigenbasis}, as well as the fact that $\boldsymbol{\sigma}_x^2 = \textbf{I}_2$, we obtain
\begin{equation}
    \textbf{q}^n = 
    \begin{cases} q^n_0
    \begin{pmatrix}
          & \textbf{U}_{\rm CR}^\dagger\boldsymbol{\sigma}_x\textbf{U}_{\rm JC}&\\
         \textbf{U}_{\rm JC}^\dagger\boldsymbol{\sigma}_x\textbf{U}_{\rm CR}& &\\
    \end{pmatrix}(n \textrm{ odd}) \\ 
     \\
    q^n_0
    \begin{pmatrix}
        \textbf{I}_2 & \\
          & \textbf{I}_2 \\
    \end{pmatrix} (n \textrm{ even}) \\
    \end{cases}
    \label{coordinate power two-state energy eigenbasis}
\end{equation}
\begin{equation}
    \boldsymbol{\mu}^n = 
    \begin{cases}
    \mu^n_0
    \begin{pmatrix}
          &\textbf{U}_{\rm CR}^\dagger\textbf{U}_{\rm JC}&\\
         \textbf{U}_{\rm JC}^\dagger\textbf{U}_{\rm CR}& &\\
    \end{pmatrix}(n \textrm{ odd}) \\ 
     \\
     \mu^n_0
    \begin{pmatrix}
        \textbf{I}_2 & \\
          & \textbf{I}_2 \\
    \end{pmatrix} (n \textrm{ even}) \\
    \end{cases}
    \label{dipole power two-state energy eigenbasis}
\end{equation}

This result presents what appears to be a contradiction. Eqs. \ref{coordinate power two-state energy eigenbasis} and \ref{dipole power two-state energy eigenbasis} seem to state that for even $n$, $\left[\hat{H}, \hat{\mu}^n \right] = \left[\hat{H}, \hat{q}^n \right] = 0$, as their matrix representations can be diagonalized in the same basis set. This would imply that the expectation values of the operators are constants of motion for even powers $n$, regardless of whether the initial condition for the cavity mode is a Fock state, coherent state, or some other state. This result disagrees with Figure \ref{q power plot} and Figure S1, which shows oscillations of the expectation values of $\hat{q}^n$ and $\hat{\mu}^n$ for even $n$ when the cavity mode initial condition is a Fock state. In addition to disagreeing with the results of our simulations, the aforementioned commutation relation can also be proven false algebraically using the definitions of these operators given in Eq. \ref{two level quantum rabi} and the subsequent text. 

The resolution of this contradiction comes when we realize that $\left[\hat{H}, \hat{\mu}^n \right] = \left[\hat{H}, \hat{q}^n \right] = 0$ for even $n$ only if we are treating both the light and matter subsystems as two-level systems, i.e., the relation is only true within the specific subspace of the joint light--matter Hilbert space considered thus far. To see how this comes about, we consider the commutator $\left[\hat{H}, \hat{\mu}^2 \right]$, which we can evaluate:
\begin{equation}
\begin{split}
    \left[\hat{H}, \hat{\mu}^2 \right] &= \left[\omega\hat{b}^\dagger\hat{b}, \, \mu_0^2\left(\hat{b}^\dagger\hat{b}+ \hat{b}\hat{b}^\dagger  +\hat{b}\hat{b} + \hat{b}^\dagger\hat{b}^\dagger\right) \right] \\
    &= \left[\omega\hat{b}^\dagger\hat{b}, \, \mu_0^2\left(\hat{b}^\dagger\hat{b}+ \left(1+\hat{b}^\dagger\hat{b}\right)  +\hat{b}\hat{b} + \hat{b}^\dagger\hat{b}^\dagger\right) \right] \\
    &= \left[\omega\hat{b}^\dagger\hat{b}, \, \mu_0^2\left(\hat{b}\hat{b} + \hat{b}^\dagger\hat{b}^\dagger\right) \right] \\
    &= 2\omega\mu_0^2(\hat{b}^\dagger\hat{b}^\dagger - \hat{b}\hat{b})
\end{split}
\label{commutator evaluation}
\end{equation}
At the second equality, we have used the commutation relation for bosonic operators $\left[\hat{b}, \hat{b}^\dagger\right] = 1$. Eq. \ref{commutator evaluation} is clearly nonzero in general, but if the molecular subsystem is taken to be a two-level system, then the double excitation (deexcitation) operator $\hat{b}^\dagger\hat{b}^\dagger$ ($\hat{b}\hat{b}$) returns zero if it acts on any state ket. These operators can therefore be taken to be zero, and we then obtain $\left[\hat{H}, \hat{\mu}^2 \right] = 0$. The same result can be shown to hold for $\left[\hat{H}, \hat{q}^2 \right]$ simply by taking the cavity mode to be a two-level system and making the replacement $\hat{b} \rightarrow \hat{a}$, recognizing that $\hat{a}$ satisfies $\left[\hat{a}, \hat{a}^\dagger\right] = 1$. This result also holds if the bosonic operators $\hat{b}$ and $\hat{b}^\dagger$ describing the molecular system are replaced by fermionic operators; the proof is given in Section 3 of the Supplementary Material. 

The above result makes it clear that we need to determine the structure of $\textbf{q}^n$ and $\boldsymbol{\mu}^n$ in a larger energy eigenbasis, i.e., we must increase the dimensionality of the subsystems. To expand the eigenbasis, we will first consider the structures of the matrices $\textbf{H}$, $\textbf{q}$, and $\boldsymbol{\mu}$ in an extended dressed-state basis. We will then convert $\textbf{q}$ and $\boldsymbol{\mu}$ into the corresponding energy eigenbasis that diagonalizes $\textbf{H}$. Finally, we will  take powers of these matrices to prove the even-odd power rule in our model.

We first prove that the block-diagonal structure of $\textbf{H}$ as a direct sum of counterrotating and Jaynes-Cummings blocks, as shown in Eq. \ref{two level quantum rabi matrix}, is a general result that holds for a dressed-state basis of any size. We start by recognizing that all states in the counterrotating block must in some sense be ``connected" to the state $\ket{00}$ through the operation of the light--matter coupling operator $\hat{V}$. For notational simplicity, we will use the subscript $\rm D$ in lieu of specifying two quantum numbers for the field and molecular degrees of freedom. Formally, two dressed states $\ket{i_{\rm D}}$ and $\ket{f_{\rm D}}$ are connected if $\ket{f_{\rm D}}$ is a member of a set $\{\ket{j_{\rm D}}\}$ such that for some positive integer $m \geq 1$, we can write
\begin{equation}
    \hat{V}^m\ket{i_{\rm D}} = \sum_j c_j\ket{j_{\rm D}}
    \label{connectivity definition}
\end{equation}
where all $|c_j| > 0$. In Eq. \ref{connectivity definition}, $\hat{V}$ is given by
\begin{equation}
\hat{V} = g(\hat{a}^\dagger + \hat{a})(\hat{b}^\dagger + \hat{b}) = g(\hat{a}^\dagger\hat{b}^\dagger + \hat{a}^\dagger\hat{b} + \hat{a}\hat{b}^\dagger + \hat{a}\hat{b})
\label{coupling}
\end{equation}

\begin{figure}[H]
\centering
\includegraphics[scale=0.5]{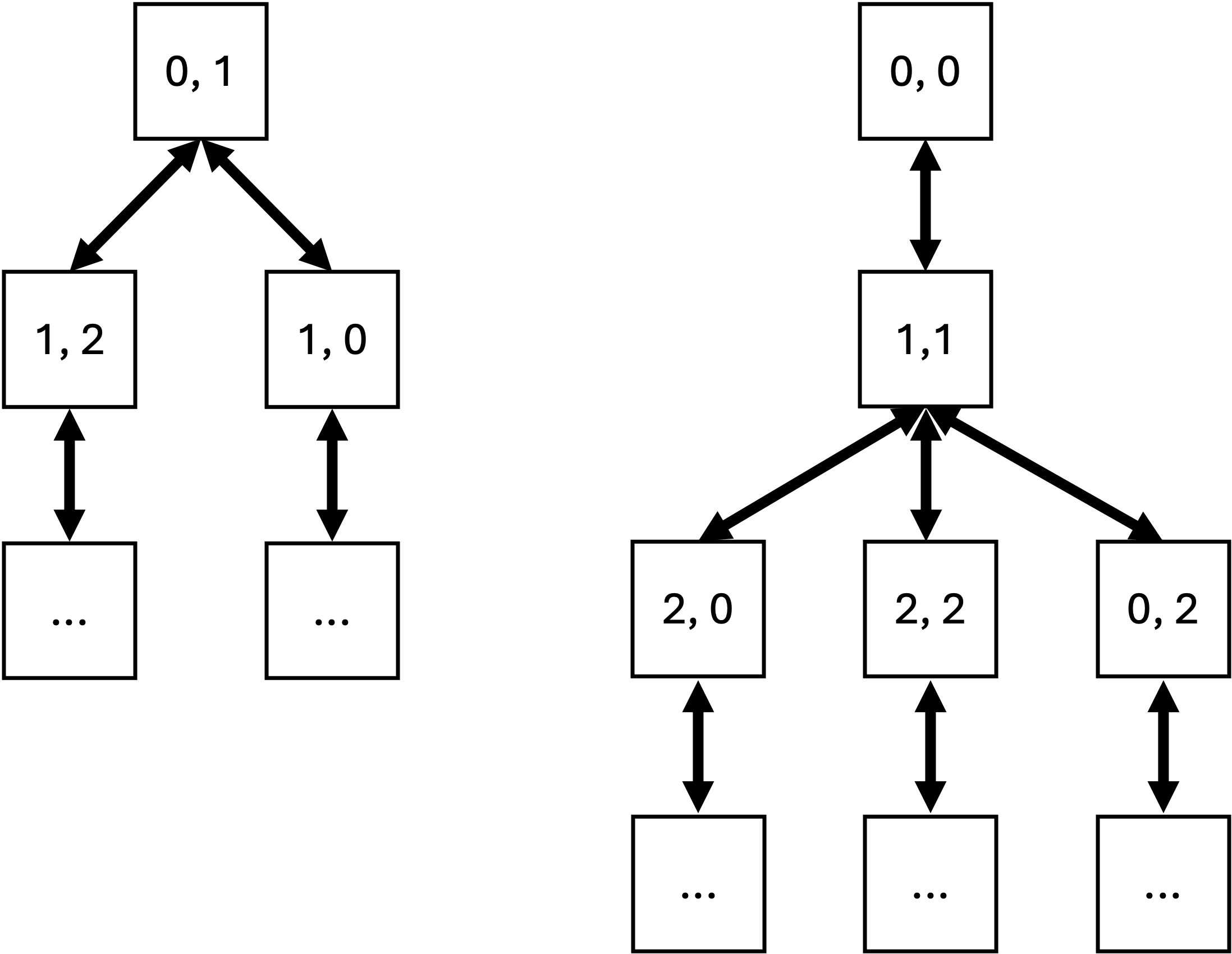}
    \caption{Structure of the trees corresponding to the Jaynes-Cummings (left) and counterrotating (right) blocks of $\textbf{H}$. Each tree is shown up to the first layer with more than one node. }
    \label{trees}
\end{figure}

Eq. \ref{coupling} defines four operations: double excitation ($\hat{a}^\dagger\hat{b}^\dagger$), excitation-deexcitation ($\hat{a}^\dagger\hat{b}$), deexcitation-excitation ($\hat{a}\hat{b}^\dagger$), and double deexcitation ($\hat{a}\hat{b}$). These operators, respectively, increase both quantum numbers $F$ and $M$ by 1; increase $F$ and decrease $M$, both by 1; decrease $F$ and increase $M$, both by 1; and decrease both quantum numbers $F$ and $M$ by 1. Any state $\ket{FM}$ is in the counterrotating block if it can be connected to the state $\ket{00}$ through any combination of these four operations. Likewise, we can also say that a state $\ket{FM}$ is in the Jaynes-Cummings block if it can be connected to the state  $\ket{01}$ through any combination of these four operations. This approach implies a tree structure of the dressed states in each of the counterrotating and Jaynes-Cummings blocks, as shown in Figure \ref{trees}. 

To prove the desired result, we must prove (a) that any state $\ket{FM}$ can be assigned to one of these two trees, and (b) that these two trees can never be connected by the set of operations given above. To prove (a), we note that for any dressed state $\ket{FM}$, repeated application of the excitation-deexcitation operator (if $F < M$) or the deexcitation-excitation operator (if $F > M$) will eventually connect to a state $\ket{GH}$ where $G = H$ or $G = H \pm 1$. Which outcome occurs will depend on whether $F - M$ is even or odd. If $F - M$ is even, then we will eventually obtain $G = H$, putting us at a state on the trunk of the counterrotating tree in Figure \ref{trees}. Repeated application of the double deexcitation operator will then eventually yield $\ket{00}$. If $F - M$ is odd, then we will eventually obtain $G = H \pm 1$. Repeated application of the double deexcitation operator will then lead to either $\ket{01}$ or $\ket{10}$; $\ket{10}$ can be connected to $\ket{01}$ by applying the deexcitation-excitation operation. Therefore, any state $\ket{FM}$ belongs to one of the two trees in Figure \ref{trees}. Furthermore, it follows from this analysis that the two trees can never be connected. The states $\ket{FM}$ in the Jaynes-Cummings tree satisfy the criterion that $F-M$ is odd, whereas the states $\ket{FM}$ in the counterrotating tree satisfy the criterion that $F-M$ is even. None of the four operations given above can change the parity of the difference $F-M$, i.e., they cannot connect a state where $F-M$ is even to one where it is odd and vice versa. Hence, the two trees cannot be connected, and the proof is complete. 

We have established that the Hamiltonian can be written in the form 
\begin{equation}
    \textbf{H} = 
    \begin{pmatrix}
        \textbf{H}_{\rm CR} & \\
         & \textbf{H}_{\rm JC} \\
    \end{pmatrix}
    \label{hamiltonian block structure}
\end{equation}
in any basis of dressed states $\ket{FM}$. This implies that the matrix $\textbf{U}$ that diagonalizes $\textbf{H}$ can be written as in Eq. \ref{transformation matrix}: 
\begin{equation}
    \textbf{U} = 
    \begin{pmatrix}
         \textbf{U}_{\rm CR} & \\
          & \textbf{U}_{\rm JC}\\
    \end{pmatrix}
\end{equation}
We can also show that the matrices $\textbf{q}$ and $\boldsymbol{\mu}$, when written in such a general dressed-state basis, have the form 
\begin{equation}
    \begin{pmatrix}
         & \textbf{M}^\dagger\\
          \textbf{M} & \\
    \end{pmatrix}
    \label{q mu matrix structure}
\end{equation}
To prove this, we note that the operators $\hat{q} = q_0(\hat{a} + \hat{a}^\dagger)$ and
$\hat{\mu} = \mu_0(\hat{b} + \hat{b}^\dagger)$ can only have matrix elements between states $\ket{FM}$ with opposite parity of the difference $F-M$, i.e., a pair of states where $F-M$ is odd for one state and even for the other. This is because a single action of $\hat{a}$ or $\hat{b}$, or their Hermitian conjugates $\hat{a}^\dagger$ and $\hat{b}^\dagger$, will change either $F$ or $M$ by 1, and thereby switch the parity of the difference $F-M$. The Jaynes-Cummings and counterrotating blocks all have states of the same difference parity, so any nonzero elements of $\textbf{q}$ and $\boldsymbol{\mu}$ must be in the off-diagonal blocks. The additional requirement that these matrices must be Hermitian implies the structure shown in Eq. \ref{q mu matrix structure}. 

Using the transformation matrix $\textbf{U}$ to transform a matrix of the form given in Eq. \ref{q mu matrix structure} into the energy eigenbasis does not change its off-diagonal block structure:
\begin{equation}
    \begin{pmatrix}
         & \textbf{M}^\dagger\\
          \textbf{M} & \\
    \end{pmatrix}
    \rightarrow
    \begin{pmatrix}
         & {\textbf{M}^\prime}^\dagger\\
          \textbf{M}^\prime & \\
    \end{pmatrix}
     = 
    \begin{pmatrix}
         & \textbf{U}^\dagger_{\rm CR}\textbf{M}^\dagger\textbf{U}_{\rm JC}\\
          \textbf{U}_{\rm JC}^\dagger\textbf{M}\textbf{U}_{\rm CR} & \\
    \end{pmatrix}
    \label{transformed M matrix}
\end{equation}
In the case of two two-level systems, as  treated above in Eqs. \ref{coordinate energy eigenbasis} and \ref{dipole energy eigenbasis}, $\textbf{M}^\prime = \textbf{U}_{\rm JC}^\dagger\boldsymbol{\sigma}_x\textbf{U}_{\rm CR}$ and $\textbf{U}_{\rm JC}^\dagger\textbf{U}_{\rm CR}$ for ${\bf q}$ and $
\boldsymbol{\mu}$, respectively. In this minimal basis set, $\textbf{M}^\prime$ is unitary for both operators; as discussed above, this is a consequence of that choice of basis set and is not true in general.

If we take powers of Eq. \ref{transformed M matrix}, we find that we alternate between an off-diagonal block structure for odd powers of $n$ and a block diagonal structure for even powers of $n$. Since $\left[\hat{H}, \hat{q}^n \right] \neq 0$ and $\left[\hat{H}, \hat{\mu}^n \right] \neq 0$, the blocks will in general not be diagonal, with the exception, as discussed above, of the minimal basis set where both the molecule and cavity mode subsystems are treated as two-level systems. Thus, for even powers $n$, there will be nonzero matrix elements between different Jaynes-Cummings energy eigenstates and between different counterrotating energy eigenstates. We will therefore observe oscillations of $q^n(t)$ and $\mu^n(t)$ for even $n$ when the dynamics is initialized in the superposition of Jaynes-Cummings eigenstates $\ket{10} = \frac{1}{\sqrt{2}}\left( \ket{\psi_+^{\rm JC}} - \ket{\psi_-^{\rm JC}} \right)$. For odd powers $n$, however, we will not observe oscillations because of the off-diagonal block structure of Eq. \ref{transformed M matrix}. This result is true for any situation where the two subsystems are each treated as an $m$-level subsystem, where $m > 2$. 

The frequencies of oscillation of $q^n(t)$ and $\mu^n(t)$ will be determined by diagonalization of $\textbf{H}$ in a given basis. In the next section, we will diagonalize the Hamiltonian given in Eq. \ref{two level quantum rabi} in the smallest basis required to observe oscillations of $q^2(t)$, and we will compare the frequencies of oscillation predicted by the model with the frequencies obtained from the power spectrum of $q^2(t)$ in Figure \ref{q square FT}. 

\subsection{Origin of fq-RT-NEO Spectral Features}
Our previous analysis will now help us understand the origins of the spectral features in the power spectrum of $q^2(t)$ shown in Figure \ref{q square FT} above. To reduce the dimensionality of our model to a minimum while still capturing all major qualitative features, we will treat the molecule as a two-level system and the cavity mode as a three-level system. As established in the previous section, treating the cavity mode as a three-level system is the minimum basis set size required to obtain nonzero matrix elements of $\hat{q}^2$ between Jaynes-Cummings energy eigenstates, and therefore to observe oscillations of $q^2(t)$ when the joint light--matter system is initialized in the dressed state $\ket{10} = \frac{1}{\sqrt{2}}\left( \ket{\psi_+^{\rm JC}} - \ket{\psi_-^{\rm JC}} \right)$. We will therefore represent the Hamiltonian in Eq. \ref{two level quantum rabi} in the basis of six dressed states: $\ket{00}, \ket{11}, \ket{20}, \ket{01}, \ket{10}$, and $\ket{21}$. As proven in the previous section, the matrix representation of Eq. \ref{two level quantum rabi} in this basis is block diagonal and factors into Jaynes-Cummings and counterrotating blocks, analogous to Eq. \ref{two level quantum rabi matrix}. 

Within this matrix, we will restrict our attention to the Jaynes-Cummings block, which contains the dressed state $\ket{10}$ corresponding to the fq-RT-NEO initial condition. This block is the matrix representation of the Hamiltonian in Eq. \ref{two level quantum rabi matrix} in the basis $\ket{01}, \ket{10}$, and $\ket{21}$, and is given by
\begin{equation}
    \textbf{H}_{\rm JC} = 
    \begin{pmatrix}
        \omega & g & \\
        g & \omega & \sqrt{2}g \\
         & \sqrt{2}g & 3\omega \\
    \end{pmatrix}
    \label{three level JC block}
\end{equation}
To determine the possible frequencies of oscillation of $q^2(t)$, we need to determine the differences between the energy eigenvalues of $\textbf{H}_{\rm JC}$. We will proceed via perturbation theory by taking the light--matter coupling matrix elements $g$ to be the perturbation and the three dressed states to be the unperturbed states. The two-fold degeneracy in the upper-left block spanned by $\ket{01}$ and $\ket{10}$ means that we first need to diagonalize the perturbation in that subspace, thereby providing the first-order perturbative correction. We can achieve this by transforming with the matrix
\begin{equation}
    \textbf{U}_{\rm JC} = \frac{1}{\sqrt{2}}
    \begin{pmatrix}
        1 & 1 & \\
        1 & -1 & \\
         & & \sqrt{2} \\
    \end{pmatrix}
    \label{three level transform}
\end{equation}
This matrix will transform $\textbf{H}_{\rm JC}$ into the zeroth-order basis $\ket{\rm UP} = \frac{1}{\sqrt{2}}(\ket{01} + \ket{10})$, $\ket{\rm LP}= \frac{1}{\sqrt{2}}(\ket{01} - \ket{10})$, and $\ket{21}$. $\ket{\rm LP}$ and $\ket{\rm UP}$ are the lower and upper polariton states that would be obtained by solving the Jaynes-Cummings model for two coupled two-level systems. Transformation of $\textbf{H}_{\rm JC}$ yields 
\begin{equation}
    \textbf{H}_{\rm JC} \rightarrow \textbf{U}_{\rm JC}^\dagger\textbf{H}_{\rm JC}\textbf{U}_{\rm JC} = 
    \begin{pmatrix}
        \omega+g &  & g\\
         & \omega-g & g\\
         g & g & 3\omega \\
    \end{pmatrix}
\end{equation}

Corrected to second order, the energies of the states $\ket{\rm LP}$, $\ket{\rm UP}$, and $\ket{21}$ are
\begin{equation}
    E_{\rm LP} = \omega - g - \frac{g^2}{2\omega + g}
    \label{LP corrected energy}
\end{equation}
\begin{equation}
    E_{\rm UP} = \omega + g - \frac{g^2}{2\omega - g}
    \label{UP corrected energy}
\end{equation}
\begin{equation}
    E_{\rm 21} = 3\omega + \frac{g^2}{2\omega + g} + \frac{g^2}{2\omega - g}
    \label{21 corrected energy}
\end{equation}
Note that the last two terms in Eq. \ref{21 corrected energy} are approximately equal for $g << \omega$. We will thus refer to them both as $A = \frac{g^2}{2\omega + g} \approx \frac{g^2}{2\omega - g}$ in Eqs. \ref{LP corrected energy} - \ref{21 corrected energy}. 
We can now compute the differences between the energy eigenvalues:
\begin{equation}
\begin{split}
    E_{\rm UP} - E_{\rm LP} = 2g = \Omega_{\rm R} \\ 
    E_{\rm 21} - E_{\rm UP} = 2\omega - g + 3A\\
    E_{\rm 21} - E_{\rm LP} = 2\omega + g + 3A
\end{split}
\label{three level differences}
\end{equation}
Eq. \ref{three level differences} indicates that within this three-state model, if we assume that $\hat{q}^2$ has nonzero matrix elements between all three eigenstates of $\textbf{H}_{\rm JC}$, which will be checked below for this model, then the power spectrum of $q^2(t)$ should show a peak at $\Omega_{\rm R} = 2g$ as well as a pair of peaks centered at approximately $2\omega$ with a splitting of $\Omega_{\rm R} = 2g$. No features appear in the region of $\omega$. 

These findings are in excellent qualitative agreement with the data shown in Figure \ref{q square FT}. Our sense of the quality of this agreement can be further enhanced by developing a comparison of the relative sizes of $2\omega$ and $A$. To obtain a rough sense of the magnitude of $A$ relative to $\omega$ and $g$, we can expand it in powers of the small quantity $g/2\omega$ to find that $A \approx g \left(\frac{g}{2\omega}\right)$. 
This result suggests that $A$ is likely multiple orders of magnitude smaller than $2\omega$, further justifying our interpretation that the pair of peaks in our three-level model will be centered very nearly at $2\omega$. If we compare our model to the fq-RT-NEO results in Figure \ref{q square FT}a, we can identify $g = \Omega_{\rm R}/2 \approx 283$ cm\textsuperscript{-1}. Given that $\omega = 2803$ cm\textsuperscript{-1}, we then find $A \approx g \left(g/2\omega \right) = 14$ cm\textsuperscript{-1}, which is more than two orders of magnitude smaller than $2\omega = 5606$ cm\textsuperscript{-1}. We acknowledge that comparison of our model results to the fq-RT-NEO results at an energy scale as small as 14 cm\textsuperscript{-1} is unlikely to succeed due to our model's simplicity. The approximations underlying our model include its low dimensionality, neglect of permanent dipole moments and transition dipoles between non-consecutive energy eigenstates, and absence of anharmonicity in the proton vibrational mode that would be captured by RT-NEO\cite{hammes-schiffer_nuclearelectronic_2021}, among other limitations. However, we are encouraged by the general qualitative agreement between this model and our fq-RT-NEO results.    

As mentioned above, we will end by checking that there are nonzero matrix elements between all of the eigenstates of $\textbf{H}_{\rm JC}$. We can write the matrix $\textbf{q}^2$ in the dressed-state basis and then transform it using $\textbf{U}_{\rm JC}$. In the dressed-state basis, we have
\begin{equation}
    \textbf{q}^2 = q_0^2
    \begin{pmatrix}
        1 &  & \sqrt{2}\\
         & 3 & \\
        \sqrt2 & & 2 \\
    \end{pmatrix}
\end{equation}
Transformation with $\textbf{U}_{\rm JC}$ then yields
\begin{equation}
    \textbf{q}^2 \rightarrow \textbf{U}_{\rm JC}^\dagger\textbf{q}^2\textbf{U}_{\rm JC} = q_0^2
    \begin{pmatrix}
        2 & -1 & 1\\
        -1 & 2 & 1\\
         1 &1 & 2 \\
    \end{pmatrix}
\label{zeroth order q square}
\end{equation}
This matrix gives $\textbf{q}^2$ in the zeroth-order basis defined above. To obtain matrix elements of $\textbf{q}^2$ in a more exact energy eigenbasis, we could perturbatively correct the zeroth-order states and further transform $\textbf{q}^2$ into this corrected basis. However, as shown in the discussion above, reasonable qualitative agreement between this three-level model and the fq-RT-NEO results can be obtained, even if perturbative corrections to the zeroth-order states are neglected. Therefore, Eq. \ref{zeroth order q square} is considered to be a good approximation to the matrix representation of $\hat{q}^2$ in the exact eigenbasis of $\textbf{H}_{\rm JC}$. Given that there are nonzero matrix elements between all three unique pairs of eigenstates of $\textbf{H}_{\rm JC}$, we can confirm that a power spectrum of our model treatment of $q^2(t)$ will have three peaks, one around $2g = 
\Omega_{\rm R}$, and the other two centered around $2\omega$ and split by $\Omega_{\rm R}$. No features will be observed in the region around $\omega$. This behavior is in excellent agreement with the fq-RT-NEO results shown in Figure \ref{q square FT}.

Finally, we point out the good qualitative agreement between the spectral features predicted by Eq. \ref{three level differences} and the features in the power spectrum of $\mu_{{\rm n}, x}^2(t)$ shown in Figure S2 of the Supporting Information. This agreement is intuitively reasonable if we assume that the dipole moment matrix $\boldsymbol{\mu}$ is similar in structure to the matrix $\boldsymbol{q}$, which we have analyzed in this section. The redshift of the center of the pair of peaks relative to $2\omega$, which also occurs to a lesser degree in the power spectrum of $q^2(t)$ shown in Figure \ref{q square FT}, likely reflects phenomena that cannot be captured by the simple three-level model employed here. 

\section{Conclusions}
In this work, we used our previously developed mfq-RT-NEO and fq-RT-NEO methods to search for unique behavior in real-time polaritonic dynamics obtained when the cavity mode is initialized in a Fock state and an HCN molecule is initialized in its ground state. Such an initial condition corresponds to the picture of coherent energy transfer in polaritonic systems obtained from model treatments. Under this initial condition, the mfq-RT-NEO dynamics show no oscillation of the expectation values of the cavity mode coordinate and molecular quantum nuclear dipole moment operators, suggesting that coherent energy exchange and the associated polariton formation between the molecule and cavity mode do not occur at this level of theory. Our analysis shows that under these initial conditions, the mfq-RT-NEO method predicts no time evolution of the molecular and cavity mode subsystems. Such behavior illustrates the limitations of the mean-field quantum method, as well as classical cavity mode approaches, which do not allow an initial cavity mode Fock state to produce polaritonic states.

Under the same initial conditions, fq-RT-NEO dynamics also show no oscillations of the expectation values of the cavity mode coordinate and molecular quantum nuclear dipole moment operators. In this full-quantum treatment, however, other evidence of polariton formation is observed. The von Neumann entropy $S(t)$, which is a measure of the quantum entanglement between the molecule and the cavity mode, oscillates significantly, reaching a maximum observed value of nearly 50\% of the theoretical maximum value that could be observed with the simulation parameters used. Such light--matter entanglement implies polariton formation caused by coherent energy exchange between the molecule and the cavity mode. Interestingly, the expectation values of even powers of the coordinate and dipole moment operators oscillate, implicating polariton formation, whereas the expectation values of odd powers of these operators remain constant throughout the dynamics. The power spectrum of the expectation value of the square of the cavity mode coordinate operator reveals three major peaks. A single peak appears in the lower-energy region of the spectrum and can be interpreted as the Rabi splitting $\Omega_{\rm R}$, while two other peaks are centered at approximately twice the cavity mode frequency and split by $\Omega_{\rm R}$. All these observations can be rationalized by comparison to the quantum Rabi model. These models also help understand what conditions are required to observe the canonical feature of polariton formation, namely, a bare molecular transition peak splitting into two polariton peaks separated by a Rabi splitting.

These findings suggest that quantum electrodynamics may demonstrate new phenomena that are inaccessible to classical electrodynamics. This analysis also highlights the limitations of the traditional Jaynes-Cummings picture of polariton formation, where both subsystems are treated as two-level systems within the rotating wave approximation. Moving beyond the rotating wave approximation and considering a larger basis set for at least one of the subsystems can be important for even a qualitative understanding of molecular polariton dynamics. Thus, these findings justify the development and use of first-principles dynamics methods in the investigation of polariton chemistry. Future work will build upon these insights in an effort to achieve an atomistic-level understanding of the chemistry of molecular polaritonic systems. 

\newpage
%%%%%%%%%%%%%%%%%%%%%%%%%%%%%%%%%%%%%%%%%%%%%%%%%%%%%%%%%%%%%%%%%%%%%
%% The "Acknowledgement" section can be given in all manuscript
%% classes.  This should be given within the "acknowledgement"
%% environment, which will make the correct section or running title.
%%%%%%%%%%%%%%%%%%%%%%%%%%%%%%%%%%%%%%%%%%%%%%%%%%%%%%%%%%%%%%%%%%%%%

\section*{Supplementary Material}
The supplementary material contains a representative Q-Chem input file and molecular geometry for the reported results, fq-RT-NEO results for $\mu_{{\rm n}, x}^n(t)$ for $n = 1,...8$, the power spectrum of $\mu_{{\rm n}, x}^2(t)$, and the proof of the commutator algebra corresponding to Eq. 35 for fermionic operators.

\begin{acknowledgements}
This material is based upon work supported by the Air Force Office of Scientific Research under AFOSR Award No. FA9550-24-1-0347. 
We thank Marissa Weichman, Scott Garner, Jonathan Fetherolf, Tim Duong, Arghadip Koner, Joel-Yuen Zhou, Chiara Aieta, Eno Paenurk, and Joseph Dickinson for helpful discussions.

\end{acknowledgements}

\section*{Data Availability Statement}
The data that support the findings of this study will be openly available in Zenodo.

%\section*{Author Declarations}
\subsection*{Conflict of Interest}
The authors have no conflicts to disclose.
%\subsection*{Author Contributions}

%%%%%%%%%%%%%%%%%%%%%%%%%%%%%%%%%%%%%%%%%%%%%%%%%%%%%%%%%%%%%%%%%%%%%
%% The same is true for Supporting Information, which should use the
%% suppinfo environment.
%%%%%%%%%%%%%%%%%%%%%%%%%%%%%%%%%%%%%%%%%%%%%%%%%%%%%%%%%%%%%%%%%%%%%

%%%%%%%%%%%%%%%%%%%%%%%%%%%%%%%%%%%%%%%%%%%%%%%%%%%%%%%%%%%%%%%%%%%%%
%% The appropriate \bibliography command should be placed here.
%% Notice that the class file automatically sets \bibliographystyle
%% and also names the section correctly.
%%%%%%%%%%%%%%%%%%%%%%%%%%%%%%%%%%%%%%%%%%%%%%%%%%%%%%%%%%%%%%%%%%%%%
%\bibliographystyle{apsrev4-2}

\bibliography{references}

\end{document}

	%%%%%%%%%% Merge with supplemental materials %%%%%%%%%%
	%%%%%%%%%% Prefix a "S" to all equations, figures, tables and reset the counter %%%%%%%%%%
	\setcounter{equation}{0}
	\setcounter{figure}{0}
	\setcounter{table}{0}
	%\makeatletter
	\renewcommand{\theequation}{S\arabic{equation}}
	\renewcommand{\thefigure}{S\arabic{figure}}
	\renewcommand{\bibnumfmt}[1]{[S#1]}
	\renewcommand{\citenumfont}[1]{S#1}
	\renewcommand{\thepage}{S\arabic{page}}
	
	\maketitle
	
	\newpage

	%%%%%%%%%% Prefix a "S" to all equations, figures, tables and reset the counter %%%%%%%%%%

\tableofcontents

\newpage

\section{Q-Chem Input File for HCN with fq-RT-NEO}

	\RecustomVerbatimCommand{\VerbatimInput}{VerbatimInput}%
	{fontsize=\footnotesize,
		frame=lines,  % top and bottom rule only
		framesep=2em, % separation between frame and text
		label=\fbox{HCN.in\_fq for Fig. 5},
		labelposition=topline,
		commandchars=\|\(\), % escape character and argument delimiters for
		% commands within the verbatim
		commentchar=*        % comment character
	}
	
\begin{Verbatim}
$molecule
0 1
C 0.0 0.0 -0.5026771429
N 0.0 0.0  0.6555628571
H 0.0 0.0 -1.5728771429 
$end

$rem
sym_ignore = 1
input_bohr = false
method = b3lyp
neo = true
neo_epc = epc172
basis = cc-pvdz
SCF_ALGORITHM = GDM
thresh = 14
s2thresh = 12
SCF_CONVERGENCE = 11
NEO_N_SCF_CONVERGENCE = 11
MAX_SCF_CYCLES = 500
NEO_PURECART = 1111
NEO_E_CONV = 12
MEM_TOTAL = 7000
NEO_VPP = 0
NEO_BASIS_LIN_DEP_THRESH = 8 
$end

$neo_basis
H    3
S    1    1.000000
   2.828400 1.0
S    1    1.000000
   4.0 1.0
S    1    1.000000
   5.6569 1.0
S    1    1.000000
   8.0 1.0
S    1    1.000000
   11.3137 1.0
S    1    1.000000
   16.0 1.0
S    1    1.000000
   22.6274 1.0
S    1    1.000000
   32.0 1.0
P    1    1.000000
   2.828400 1.0
P    1    1.000000
   4.0 1.0
P    1    1.000000
   5.6569 1.0
P    1    1.000000
   8.0 1.0
P    1    1.000000
   11.3137 1.0
P    1    1.000000
   16.0 1.0
P    1    1.000000
   22.6274 1.0
P    1    1.000000
   32.0 1.0
D    1    1.000000
   2.828400 1.0
D    1    1.000000
   4.0 1.0
D    1    1.000000
   5.6569 1.0
D    1    1.000000
   8.0 1.0
D    1    1.000000
   11.3137 1.0
D    1    1.000000
   16.0 1.0
D    1    1.000000
   22.6274 1.0
D    1    1.000000
   32.0 1.0
****
$end

$neo_tdks
dt 0.04
photon_type 2
is_coherent false 1
photon_basis_size 4
maxiter 200000
field_type delta
field_amp 0.001
in_cavity true 0.3475 0 8e-4 1e8
rt_thresh 4
$end
\end{Verbatim}
    
The above input file generates the fq-RT-NEO data given in the main text. The \textit{\$neo\_tdks} section controls the RT-NEO dynamics. ``\textit{photon\_type}" determines the level of theory used and is set to 2 for the full-quantum fq-RT-NEO method. ``\textit{is\_coherent}" initializes the cavity mode in a coherent state and is set to ``false" for all calculations in this work. ``1" specifies that one quantum of energy is to be added to the initial condition of the cavity mode. ``\textit{in\_cavity true 0.3475 0 8e-4 1e8}" denotes coupling of the molecule to a single-mode cavity with a mode frequency of 0.3475 eV, polarization direction $x$ (0 for $x$, 1 for $y$, 2 for $z$), light--matter coupling $\varepsilon = 8 \times 10\textsuperscript{-4}$ a.u., and cavity lifetime $1/\gamma_{\rm c} = 1 \times 10^8$ a.u. The large cavity lifetime indicates that cavity loss is negligible in these simulations. Finally, ``\textit{rt\_thresh}" sets the threshold of the real-time predictor-corrector algorithm to 10\textsuperscript{-4}; this is the \textit{eps} parameter in Algorithm 1 of Ref. \citenum{de_santis_pyberthart_2020}. All other parameters are assumed to be self-explanatory. 

\newpage

\section{fq-RT-NEO Results for $\boldsymbol{\mu_{{\rm n}, x}^n(t), \, n = 1,...8}$}

\begin{figure}[H]
\centering
\includegraphics[scale=1.0, trim=2 2 2 2, clip]{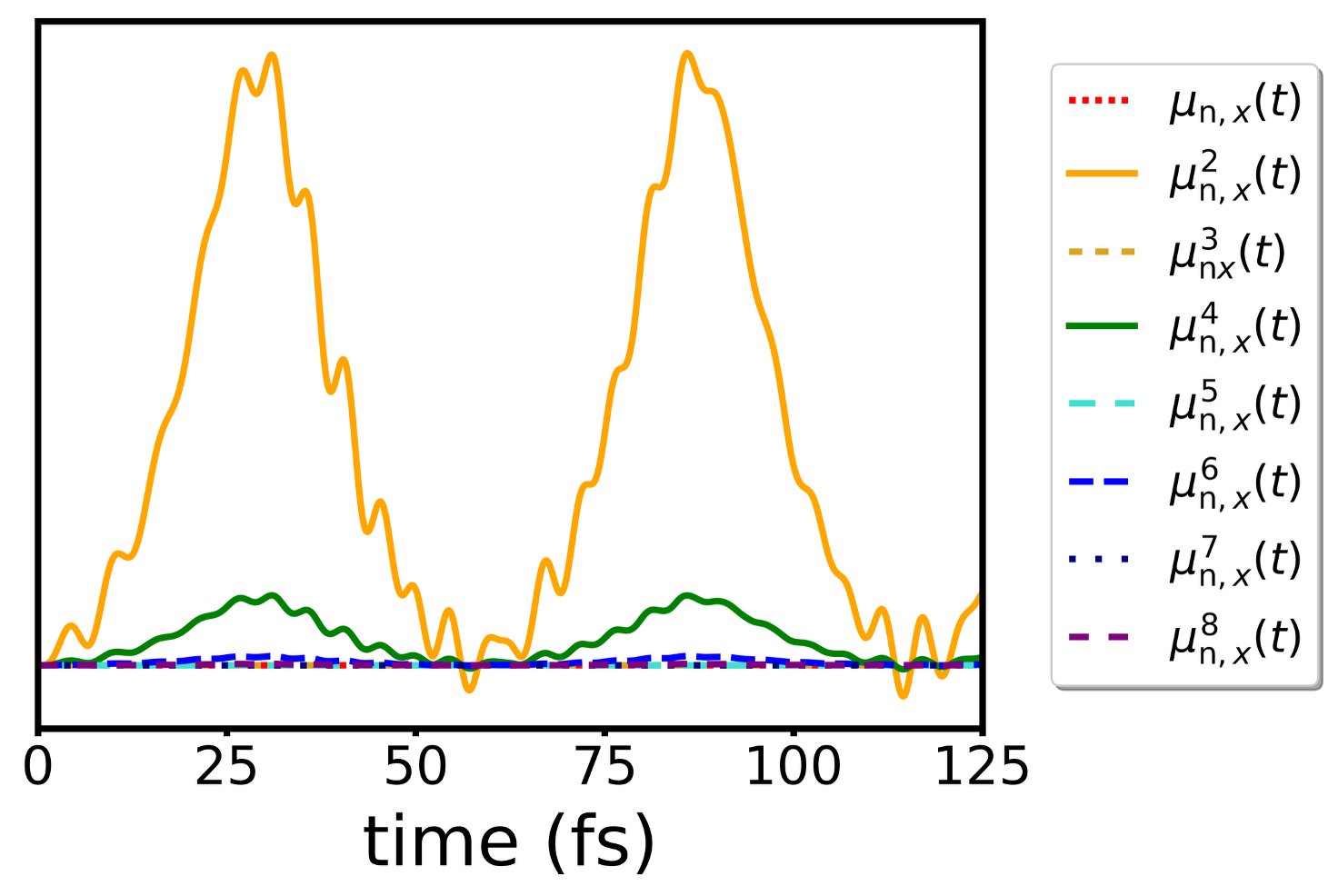}
    \caption{fq-RT-NEO dynamics of $\mu^n_{\rm n, \lambda}(t)$ for $n = 1,...8$ for HCN. No units are given on the $y$-axis because these observables have different units.}
    \label{dipole powers}
\end{figure}
Figure \ref{dipole powers} shows the fq-RT-NEO dynamics of the expectation values of $\mu^n_{\rm n, \lambda}(t) \equiv \textrm{Tr}[\textbf{P}_{\rm n}(t)\boldsymbol{\mu}^n_{\rm n, \lambda}] - \textrm{Tr}[\textbf{P}_{\rm n}(0)\boldsymbol{\mu}^n_{\rm n, \lambda}]$ for $n = 1 ... 8$. The expectation values for $\mu^2_{\rm n, \lambda}(t)$ and $\mu^4_{\rm n, \lambda}(t)$ are clearly nonzero, oscillating with a slow period of $\sim$ 59 fs and a fast oscillation period of $\sim$ 6 fs. The latter period roughly corresponds to half the period of the oscillations of the free cavity mode, $2\pi/\omega_{\rm c}$ = 12 fs. 
The expectation values for odd $n$ are all negligible, consistent with the arguments for the model system discussed in the main text. 

\newpage

\section{Fourier Transform of $\boldsymbol{\mu_{{\rm n}, x}^2(t)}$}

\begin{figure}[H]
\centering
\includegraphics[scale=1.0, trim=2 2 2 2, clip]{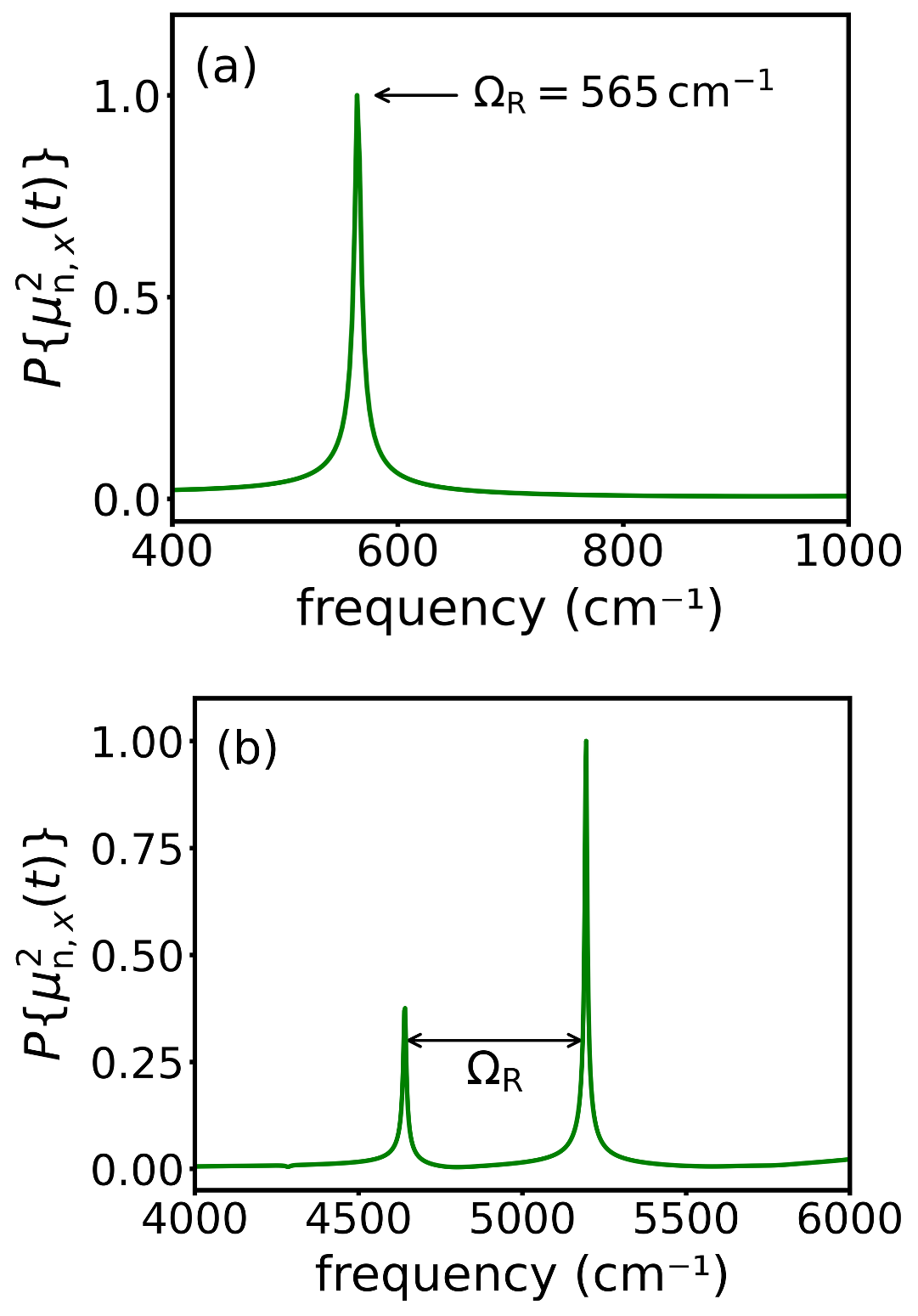}
    \caption{Fourier transform of $\mu_{\rm n, \lambda}^2(t)$ obtained with fq-RT-NEO dynamics applied to HCN for the (a) 400 -- 1000 cm\textsuperscript{-1} region and (b) 4000 -- 6000 cm\textsuperscript{-1} region. In both spectra, the largest signal amplitude has been scaled to a value of 1.}
    \label{mu square FT}
\end{figure}
Figure \ref{mu square FT}a shows the Fourier transform of $\mu_{\rm n, \lambda}^2(t)$ in the region of 400 -- 1000 cm\textsuperscript{-1}. A single peak at 565 cm\textsuperscript{-1} matches the peak interpreted as $\Omega_{\rm R}$ and shown in Figure 2a in the main text. Figure \ref{mu square FT}b shows the Fourier transform of $\mu_{\rm n, \lambda}^2(t)$ in the region of 4000 -- 6000 cm\textsuperscript{-1}. A pair of peaks is separated by almost exactly $\Omega_{\rm R}$ and centered at 4918 cm\textsuperscript{-1}, which is redshifted from $2\omega$ = 5606 cm\textsuperscript{-1} by $\sim$ 12\%. These results are in qualitative agreement with the three-level model presented at the end of the Discussion in the main text. 
%The redshift of the center of the pair of peaks, which also occurs in the Fourier transform of $q^2(t)$ shown in Fig. 3b in the main text, probably reflects the limited ability of the model to replicate the phenomena of the fq-RT-NEO result. 

\newpage

\section{Evaluation of $\boldsymbol{\left[\hat{H}, \hat{\mu}^2\right]}$ with Fermionic Operators}
We will evaluate $\left[\hat{H}, \hat{\mu}^2\right]$, where the dipole operator $\hat{\mu}$ is given by
\begin{equation}
    \hat{\mu} = \mu_0(\hat{c}^\dagger + \hat{c})
    \label{fermionic dipole}
\end{equation}
and the Hamiltonian $\hat{H}$ is given by
\begin{equation}
\begin{split}
    \hat{H} &= \hat{H}_{\rm F} + \hat{H}_{\rm M} + \varepsilon\hat{q}\hat{\mu}\\
    &= \hat{H}_{\rm F} + \hat{H}_{\rm M} + \varepsilon q_0(\hat{a}^\dagger + \hat{a})\mu_0(\hat{c}^\dagger + \hat{c}) \\
    &= \omega\hat{a}^\dagger\hat{a} + \omega\hat{c}^\dagger\hat{c} + g(\hat{a}^\dagger + \hat{a})(\hat{c}^\dagger + \hat{c}).
\end{split}
\label{fermionic two level quantum rabi}
\end{equation}

Eq. \ref{fermionic two level quantum rabi} is identical to Eq. 17 in the main text, except that the bosonic annihilation (creation) operator $\hat{b}$ ($\hat{b}^\dagger$) has been replaced by the fermionic annihilation (creation) operator $\hat{c}$ ($\hat{c}^\dagger$). To evaluate the commutator, we first invoke the Jacobi Identity
\begin{equation}
    \left[\hat{H}, \hat{\mu}^2\right] = \left[\hat{H}, \hat{\mu}\right]\hat{\mu} + \hat{\mu}\left[\hat{H}, \hat{\mu}\right].
    \label{jacobi}
\end{equation}
We then evaluate $\left[\hat{H}, \hat{\mu}\right]$ using the fermionic commutation relation $\left[\hat{c}, \hat{c}^\dagger\right] = 1 - 2\hat{c}^\dagger \hat{c}$, as well as Eqs. \ref{fermionic dipole} and \ref{fermionic two level quantum rabi}:
\begin{equation}
\begin{split}
    \left[\hat{H}, \hat{\mu}\right] &= \omega\mu_0\left[\hat{c}^\dagger\hat{c}, \left(\hat{c}^\dagger+\hat{c}\right)\right] \\ 
    &= \omega\mu_0 \left( \hat{c}^\dagger \hat{c} \hat{c}^\dagger - \hat{c}^\dagger \hat{c}^\dagger \hat{c} + \hat{c}^\dagger \hat{c} \hat{c} - \hat{c} \hat{c}^\dagger \hat{c} \right) \\
    &= \omega \mu_0 \left( \hat{c}^\dagger \left(1 - 2\hat{c}^\dagger \hat{c} 
    \right) - \left(1 - 2\hat{c}^\dagger \hat{c} 
    \right) \hat{c}\right) \\
    &= \omega \mu_0 \left( \hat{c}^\dagger - 2\hat{c}^\dagger\hat{c}^\dagger \hat{c}  - \hat{c} + 2\hat{c}^\dagger \hat{c} \hat{c} \right) .
\end{split}
\end{equation}
In a minimal two-level basis, terms with two consecutive creation or annihilation operators are equal to the zero operator and can be eliminated, leading to
\begin{equation}
    \left[\hat{H}, \hat{\mu}\right] = \omega  \mu_0 \left(\hat{c}^\dagger - \hat{c} \right).
    \label{fermion hamiltonian dipole commutator}
\end{equation}
Substituting Eq. \ref{fermion hamiltonian dipole commutator} into Eq. \ref{jacobi}, we obtain the final result:
\begin{equation}
    \left[\hat{H}, \hat{\mu}^2\right] = \omega  \mu_0^2 \left[\left(\hat{c}^\dagger - \hat{c} \right)\left(\hat{c}^\dagger + \hat{c} \right) + \left(\hat{c}^\dagger + \hat{c} \right)\left(\hat{c}^\dagger - \hat{c} \right)\right] = 2\omega\mu_0^2 \left(\hat{c}^\dagger\hat{c}^\dagger - \hat{c}\hat{c}\right).
\end{equation}
This is equivalent to Eq. 35 in the main text, with $\hat{b} \rightarrow \hat{c}$. It therefore follows that $\left[\hat{H}, \hat{\mu}^2\right] = 0$ within a minimal two-level basis for both fermionic and bosonic molecular systems. 

\newpage

\bibliography{SI}